\documentclass[12pt]{article}
\usepackage{latexsym}

\def\dalemb#1#2{{\vbox{\hrule height .#2pt
        \hbox{\vrule width.#2pt height#1pt \kern#1pt
                \vrule width.#2pt}
        \hrule height.#2pt}}}

\let\a=\alpha \let\b=\beta \let\g=\gamma \let\d=\delta \let\e=\epsilon
\let\z=\zeta  \let\th=\theta  \let\k=\kappa
\let\l=\lambda \let\m=\mu \let\n=\nu \let\x=\xi  
\let\s=\sigma \let\t=\tau    
 
      \let\G=\Gamma  \let\Th=\Theta 
\let\X=\Xi    \let\Y=\Psi
 
\let\la=\label  
  
\def\nn{\nonumber} \def\bd{\begin{document}} \def\ed{\end{document}}
\def\ds{\documentstyle} \let\fr=\frac \let\bl=\bigl \let\br=\bigr
\let\Br=\Bigr \let\Bl=\Bigl
\let\bm=\bibitem
\let\na=\nabla
\def\tU{{\widetilde U}}
\let\pa=\partial \let\ov=\overline
\def\ie{{\it i.e.\ }}
\newcommand{\be}{\begin{equation}}
\newcommand{\ee}{\end{equation}}
\def\ba{\begin{array}}
\def\ea{\end{array}}
\def\ft#1#2{{\textstyle{{\scriptstyle #1}\over {\scriptstyle #2}}}}
\def\fft#1#2{{#1 \over #2}}
\def\F#1#2{{ F_{#1}^{(#2)} }}
\def\cF#1#2{{ {\cal F}_{#1}^{(#2)} }}

\def\R{{\bf R}}
\def\sst#1{{\scriptscriptstyle #1}}
\def\oneone{\rlap 1\mkern4mu{\rm l}}
\def\e7{E_{7(+7)}}
\def\td{\tilde}
\def\wtd{\widetilde}
\def\im{{\rm i}}
\def\bog{Bogomol'nyi\ }
\newcommand{\ho}[1]{$\, ^{#1}$}
\newcommand{\hoch}[1]{$\, ^{#1}$}
\newcommand{\bea}{\begin{eqnarray}}
\newcommand{\eea}{\end{eqnarray}}
\newcommand{\ra}{\rightarrow}
\newcommand{\lra}{\longrightarrow}
\newcommand{\Lra}{\Leftrightarrow}
\newcommand{\ap}{\alpha^\prime}
\newcommand{\bp}{\tilde \beta^\prime}
\newcommand{\cB}{{\cal B}}
\newcommand{\cO}{{\cal O}}
\newcommand{\vecx}{\vec{x}}
\newcommand{\vecy}{\vec{y}}
\newcommand{\vecp}{\vec{p}}
\newcommand{\vecq}{\vec{q}}
\newcommand{\tr}{{\rm tr} }
\newcommand{\Tr}{{\rm Tr} }
\newcommand{\NP}{Nucl. Phys. }

\newcommand{\cL}{{\cal L}}
\newcommand{\cA}{{\cal A}}
\newcommand{\cD}{{\cal D}}

\def\sst#1{{\scriptscriptstyle #1}}
\def\0{{\sst{(0)}}}
\def\1{{\sst{(1)}}}
\def\2{{\sst{(2)}}}
\def\3{{\sst{(3)}}}
\def\4{{\sst{(4)}}}
\def\5{{\sst{(5)}}}
\def\6{{\sst{(6)}}}
\def\7{{\sst{(7)}}}
\def\8{{\sst{(8)}}}
\def\9{{\sst{(9)}}}
\def\ve{\varepsilon}
\def\vf{\varphi}
\def\F{\Phi}
\def\wg{\wedge}

\newcommand{\tamphys}{\it 
}

\newcommand{\auth}{AUTHORS}

\def\thb{\bar{\theta}}
\def\Thb{\bar{\Theta}}
\def\barp{\bar{p}}
\def\barq{\bar{q}}
\def\barc{\bar{c}}
\def\bard{\bar{d}}
\def\e{\epsilon}

\def \bi{\bibitem}
\def \la {\label}

\def \l {\lambda}
\def\foot{\footnote}
\def \tl  {{\tilde \l}}
\def \sql {{\sqrt \l}}
\def \adss {$AdS_5 \times S^5$\ }
\newcommand{\rf}[1]{(\ref{#1})}
\def \ov {\over}

\def\th{\theta}
\def\Th{\Theta}
\def\vth{\vartheta}
\def\btheta{{\bar\theta}}
\def\ttheta{{{\tilde\theta}}}
\def\bttheta{{{\bar\ttheta}}}
\def\vth{\vartheta}

\def\ra{\rightarrow}
\def\N{{\cal N}}
\def\F{{\cal F}}
\def\uM{\underline{M}}
\def\uN{\underline{N}}
\def\uP{\underline{P}}
\def\cc{\circ}
\def\eqv{\equiv}

\def\ni{\noindent}

\def\Ep{E^{{}^{(+)}}}
\def\Em{E^{{}^{(-)}}}

\def\Mp{M^{{}^{(+)}}}
\def\Mm{M^{{}^{(-)}}}

\def \ha{{1\ov 2}}

\def\r{\rho}

\def\Y{{\rm Y}}
\def\X{{\rm X}}
\def\tY{\tilde{\rm Y}}
\def\tX{\tilde{\rm X}}
\def\dY{\dot{\rm Y}}
\def\dX{\dot{\rm X}}

\def \J {\mathcal{J}}
\def \del {\partial}

\def\dF{\dot{F}}
\def\dG{\dot{G}}
\def\df{\dot{f}}
\def \E {{\cal E}}
\def \J {{\cal J}}

\def\ms{\mathcal{S}}
\def\mj{\mathcal{J}}
\def\soj{\fr{\ms}{\mj}}
\def \R {{\bf R}}
\def \om {\omega}
\def \bE {\bar E}
\def \x {{\cal X}}

\def \bi{\bibitem}
\def \la {\label}

\def \l {\lambda}
\def\foot{\footnote}
\def \tl  {{\tilde \l}}
\def \sql {{\sqrt \l}}
\def \adss {$AdS_5 \times S^5$\ }
\def \ov {\over}

\def \varpi {{\rm w}}

\def\thb{\bar{\theta}}
\def\Thb{\bar{\Theta}}
\def\zb{\bar{z}}
\def\psib{\bar{\psi}}
\def\barp{\bar{p}}
\def\barq{\bar{q}}
\def\barc{\bar{c}}
\def\bard{\bar{d}}
\def\e{\epsilon}
\def\wb{\bar{w}}
\def\lb{\bar{\l}}
\def\Jb{\bar{J}}
\def\Nb{\bar{N}}

\def\At{\tilde{A}}
\def\Bt{\tilde{B}}
\def\Ct{\tilde{C}}
\def\Dt{\tilde{D}}
\def\Et{\tilde{E}}
\def\Ft{\tilde{F}}
\def\Gt{\tilde{G}}
\def\Mt{\tilde{M}}
\def\at{\tilde{a}}
\def\bt{\tilde{b}}
\def\ct{\tilde{c}}
\def\dt{\tilde{d}}
\def\et{\tilde{e}}
\def\ft{\tilde{f}}
\def\gt{\tilde{g}}
\def\ola{\overleftarrow}
\def\ora{\overrightarrow}
\def\at{\tilde{\a}}

\def\ps{\rlap{\, /}\;\,p }
\def\ks{\rlap{\, /}\;\,k }

\def\gym{g_{YM}}

\def\adot{\dot{a}}
\def\bdot{\dot{b}}
\def\bpa{\bar{\pa}}

\begin{document}
\overfullrule=0pt
\parskip=2pt
\parindent=12pt
\headheight=0in \headsep=0in \topmargin=0in
\oddsidemargin=0in

\vspace{ -3cm}
\thispagestyle{empty}

\begin{center}

{\Large\bf Pure spinor computation towards open string three-loop
  }

 \vspace{.5cm} { I.Y. Park  }\\
 \vskip 0.2cm

{\it Department of Physics, Kyoto University, Kyoto
 606-8502, Japan}

\vspace{0.5cm} {\it High Energy Accelerator
 Research Organization (KEK)\\
  Tsukuba, Ibaraki 305-0801, Japan}

\vspace{0.5cm}
{\it Department of Natural and physical Sciences,
Philander Smith College\footnote{Home institute}\\
Little Rock, AR 72223, USA \\
inyongpark05@gmail.com}


\end{center}

 \vspace{0.1cm}

 \begin{abstract}
\ni Using the recent results in the pure spinor formulation, we lay
out a ground-work towards the full momentum space amplitudes of open
superstrings at three-loop. After briefly reviewing the one-loop
amplitude, we directly work out the two-loop amplitude and reproduce
the result that was obtained by a symmetry argument before. For the
three-loop, first we use the two-loop regulator as a warm-up
exercise. The result vanishes. We then employ the regulator that has
been recently proposed by Aisaka and Berkovits. It is noted that the
terms in higher power in $\fr1{\l\l}$ that render the two-loop
regulator disqualified for the three-loop do not contribute. This
with a few other indications suggests a possibility that the AB
regulator might also lead to a vanishing result. Nevertheless, we
argue that it is possible to acquire the three-loop amplitude, and
present a result that we anticipate to be the three-loop amplitude.

\end{abstract}
\newpage

\setcounter{equation}{0}
\setcounter{footnote}{0}
\setcounter{section}{0}


\section{Introduction}

D-brane physics \cite{Polchinski:1995mt,pol,Johnson:2000ch} has
brought many groundbreaking results in the recent developments in
string theory and phenomenology. In particular, it has opened up a
new route to understanding gravitational physics through gauge
degrees of freedom
\cite{Banks:1996vh,Ishibashi:1996xs,Maldacena:1997re,
Gubser:1998bc,Witten:1998qj,Aharony:1999ti}. At the heart of this
phenomenon lies the fact that a D-brane admits two different
descriptions, one as a flat hyperplane on which open strings move
with their end points attached, and the other through a solitonic
solution of closed string theory.

Most of the studies so far have been devoted to analyzing how the
gauge/gravity correspondence works and discovering more examples.
While those are meaningful tasks, an equally important task would be
to understand {\em why} there is such a correspondence. Once
unraveled, the underlying mechanism is likely to lead to a
derivation from the first principle of the correspondence in its
{\em precise} form. At the very least it will hint at a route to a
derivation. The derivation is especially important since there are
some outstanding questions. For example, are there circumstances
where it is necessary to take into account the massive modes of an
open string on top of the massless modes, the SYM? Another question
is, does the conjecture remain valid in the non-planar regime?

A close interplay between an open string and a closed string must be
responsible for the correspondence.\footnote{In relation to this ,
we find the works of \cite{Kawai:2007ek} interesting.} We believe
that the mechanism underlying the gauge/gravity duality should be
sought \cite{Park:2001bm} in the unity of degrees of freedom between
an open string and a closed string
\cite{Nielsen:1973qs,Gibbons:2000hf,Sen:2000kd}. If true, it would
imply that the physics associated with an open string would be
closely related to that of a closed string. A conjecture that may be
viewed hinging on such an interplay was put forward in
\cite{Park:2007mc} and verified in the subsequent work
\cite{Park:2008fp} at one-loop. The two-loop extension was also
discussed in \cite{Park:2009ki}. There it was demonstrated that the
open string loop divergences on a D-brane may be removed by a vertex
operator constructed out of the curved geometry of the D-brane.

As part of the endeavor to extend the conjecture to the three-loop,
we study the three-loop amplitudes of open strings. Unlike the one-
and two-loop cases, no full momentum space amplitude has been
computed at this level. Therefore it is necessary to compute the
amplitudes themselves before attempting to extend the conjecture to
the three-loop. A multi-loop computation in superstring theory is a
complicated task. The complete two-loop amplitudes have been
computed only fairly recently \cite{Zheng:2002ji,D'Hoker:2005jc}.
Since the three-loop computations will be much more complicated, it
is very important to use an effective method. To our view, the pure
spinor formulation \cite{Berkovits:2000fe}, which was developed by
using some of the ingredients in \cite{Siegel:1985xj}, may provide
such a tool. Also see
\cite{Berkovits:2002qx,Berkovits:2002ag,Berkovits:2000ph,Berkovits:2006bk,
Aisaka:2002sd,Grassi:2004xr,Policastro:2006vt,Alexandrov:2007pd,
Mafra:2009wq,Gerigk:2009va} for the further developments of the
formulation. For example, compared with the RNS formulation, it
requires far less amount of algebra to obtain the two-loop amplitude
by direct computation as we show in section 3 below.\footnote{After
we finished our computation, a work \cite{Gomez:2010ad} had appeared
that had some overlap with our two-loop computation.}

 For the three-loop order, first we use the
two-loop regulator as a warm-up exercise. The result vanishes. We
then employ the regulator that has been recently proposed by Aisaka
and Berkovits (AB). It is noted in section 3 that the terms in
higher power in $\fr1{\l\l}$ that render the two-loop regulator
disqualified for the three-loop do not contribute. This with a few
other indications suggests a possibility that the AB regulator might
also lead to a vanishing result. This might be a signal of
incompleteness of the AB regulator.\footnote{We believe that there
is a room for a better understanding of a three-loop regulator. It
deserves a separate study. We will further comment on this in the
section 4 and the conclusion. } To determine whether the entire
three-loop indeed vanishes, all of the overall numerical
coefficients of the individual terms must be kept track of. As will
be fathomed in the subsequent sections, it is an extremely tedious
task\footnote{In spite of the expected tediousness, tracking precise
numerical coefficients for an amplitude in general should be an
executable task. However, it would be more meaningful to carry out
the task for a regulator that is free of any subtlety. The task is
best left to future efforts after further progress on the regulator
issue is made. } that may require some efficient method of
organization. Our main goal is to determine the form of the
three-loop amplitude, and we argue that it is possible to acquire
the three-loop amplitude without using a precise and complete
regulator. The basic idea is to rely on a pattern that the
computations reveal. We make an observation on the pattern starting
with the two-loop case below. Observed in a few chosen cases of the
three-loop as well, the pattern is expected to persist in the higher
loop orders.

The rest of the paper is organized as follows. In this paper, the
four-point amplitude is analyzed. In the beginning of the next
section, we briefly review the one-loop case to set the notations
and convention. The two- and three- loop computations involve
lengthy algebra. We present a summary of the results with a
discussion of the pattern that plays an important role in
determining the three-loop amplitude. In section 3, we directly
calculate the two-loop amplitude. The result is expressed in terms
of the SYM multiplets and the covariant derivatives, thereby
reproducing the result that was first obtained by a symmetry
argument. The simplified measure introduced in \cite{Gomez:2009qd}
significantly reduces the amount of algebra that has be carried out.
At three-loop a complete gauge invariant regulator is not currently
available, at least not in a simple form. In section 4, we carry out
our analysis using the regulator that has recently been proposed by
Aisaka and Berkovits \cite{Aisaka:2009yp}. After carrying out
$[ds]$- and $dd$- integrations, we evaluate a few terms. We propose
an expression, eq.\rf{3loopresult}, that we anticipate to be the
correct three-loop amplitude. We end with future directions in the
conclusion where few necessary checks and/or works are listed either
to strengthen or to confirm our three-loop result.

\section{Summary of results}

Since the sections 3 and 4 on the loop-computations involve lengthy
and tedious algebras, it may be useful to summarize the results
before we embark on the computations. We start in section 2.1 by
briefly reviewing the non-minimal formulation and the one-loop
computation. This will set our notations and convention. In section
2.2, a summary of the two- and three- loop results is presented.
There we ponder on the pattern that the two- and three- loop
computations reveal. The three-loop amplitude, (\ref{3loopresultq})
(quoted from eq.(\ref{3loopresult}) in section 4), has been obtained
based on the pattern.

\subsection{review of non-minimal formalism and one-loop }
There are two versions of the pure spinor formulation:the minimal
and the non-minimal. In this paper we use the non-minimal version.
Our notations and some useful identities are summarized in
Appendices A and B. We review several ingredients of the formulation
that we will need in the calculations in sections 3 and 4. For more
details, we refer to \cite{Berkovits:2000ph,Mafra:2009wq}.

The unintegrated and the integrated vertex operators that represent
the massless modes are respectively given by
 \bea
 V&=&\l^\a A_\a\nn\\
 U&=&\pa\th^\a A_\a+\Pi^m A_m+d_\a W^\a+\fr12N^{mn}{\cal F}_{mn}
 \eea
where
 \bea
 \Pi^m=\pa X^m+\fr12 \th^\b \g^m \pa \th
 \eea
 The SYM multiplets satisfy
 \bea
 A_m&=&\fr18D_\a\g_m^{\a\b}A_\b\nn\\
 W^\b&=&\fr1{10}\g_m^{\a\b}(D_\a A^m-\pa^m A_\a)\nn\\
 {\cal F}_{mn}&=&\fr18 D_\a(\g_{mn})^\a{}_\b W^\b=\pa_{[m}A_{n]}
 \label{AF}
 \eea
and
 \bea
 &&-D_\a A_\b -D_\b A_\a+\g_{\a\b}^m A_m=0 \nn\\
 && D_\a A_m-\pa_m A_\a-\g_{m\a\b}W^\b=0\nn\\
 && -D_\a W^\b+\fr14(\g^{mn})_\a{}^\b \F_{mn}=0\nn\\
 &&\l^\a\l^\b (\g_{mn})_\b{}^\g D_\a \F^{mn}=0
 \label{fieldeqs}
 \eea
 where the 16 by 16 gamma matrices, $\g_{mn}$, satisfy
 \bea
 \eta_{mn}\g^m_{\a(\b} \g^n_{\g\d)}=0 \label{etagg}
 \eea
Using $D_\a \F^{mn}=2k^{[m}\g_{\a\b}^{n]} W^\b$ and $ -D_\a
W^\b+\fr14(\g^{mn})_\a{}^\b \F_{mn}=0$, one can show
 \bea
 &&D_\b D\a \F_{mn}=-\fr12 k_{[m}(\g_{n]}\g^{tu})_{\a\b}\F_{tu}\nn\\
 && D_\d D_\g D_\b D\a \F_{mn}=\fr14
                 k_{[m}(\g_{n]}\g^{tu})_{\a\b}
                 k_{t}(\g_{u}\g^{vw})_{\g\d}\F_{vw}
                 \label{DWeqs}
 \eea
 from which it follows
 \bea
 D_{\a_1}D_{\a_2}W^{\b_1}&=&\fr12
   k^m(\g_{mn})_{\a_2}{}^{\b_1}(\g^n)_{\a_1\g}W^\g
 \nn\\
 D_{\a_1}D_{\a_2}D_{\a_3}W_\1^{\b_1}
 &=&-\fr18 (\g^{l_1l_2})_{\a_3}{}^{\b_1}k^\1_{l_1}
  (\g_{l_2}\g^{v_1v_2})_{\a_2\a_1}\F_{v_1v_2}^\1
 \eea
 In an abelian case, one has
 \bea
 \g^m_{\a\b}\pa_m W^\b=0 \label{W-eq-abelian}
 \eea
It implies on shell
 \bea
 \g^m_{\a\b}k_m W^\b=\ks W=0 \label{W-abelian-onshell}
 \eea
which is used in the two-loop computation in section 3. The pure
spinor formulation provides algorithms for $N$-loop amplitudes. Up
to and including the three-loop, they are given by
 \bea
{\cal A}_{1-loop}&=&\int d\t <{\cal N}_1(y)\int dw \m(w)b(w)V_1(z_1)
 \int dz_2 U_2(z_2)\cdots  \int dz_N U_N(z_N)>\nn\\
 {\cal A}_{2-loop}&=&\int d\t_1d\t_2d\t_3 <{\cal N}_2(y)
 \prod_{s=1}^3\int dw_s \m(w_s)b(w_s)
 \int dz_1 U_1(z_1)\cdots  \int dz_N U_N(z_N)>\nn\\
  {\cal A}_{3-loop}&=&\int d\t_1d\cdots d\t_6 <{\cal N}_3(y)
 \prod_{s=1}^6\int dw_s \m(w_s)b(w_s)
 \int dz_1 U_1(z_1)\cdots  \int dz_N U_N(z_N)>\nn\\
\label{loopprescription}
 \eea
where the ghost operator, $b$, is given by
 \bea
 b &=& s^\a\pa \lb_\a+\fr{\lb_\a\left[2\Pi^m(\g_m d)^\a
              -N_{mn}(\g^{mn}\pa\th)^\a-J_\l\pa\th^\a
              -\pa^2\th^\a\right]}{4\lb\l}
 \eea
 \bea
 +\fr{(\lb\g^{mnp}\,r)(d\g^{mnp}d+24N_{mn}\Pi_p)}{192(\lb\l)^2}
 -\fr{(r\g_{mnp}r)(\lb\g^md)N^{np}}{16(\lb\l)^3}
 +\fr{(r\g_{mnp}r)(\lb\g^{pqr}r)N^{mn}N_{qr}}{128(\lb\l)^4}
    \nn
 \eea
 The $\N_g, (g=1,2,3)$ in (\ref{loopprescription}) is the $g$-loop
regulator. Each field has different number of zero modes: for the
bosonic fiilds,
 \[
    \begin{array}{cccccc}
        x^m   & \l^\a & w_\a & \lb_\a & \bar{w}^\a  \\
        10    & 11    & 11g  & 11     &   11g \\
   \end{array}
\]
 and for the fermionic fields,
\[
    \begin{array}{cccc}
        \th^\a  &  d_\a  & r  &  s   \\
          16    &   16g  & 11 & 11g    \\
   \end{array}
\]

 For the two- and the three- loop amplitudes, we will make
frequent use of the 32-component Fierz rearrangement lemma, which is
reviewed in Appendix B. Because of this, we embed the 16-component
notation in the 32-component notation. Let us define
 \bea
 \l_u=\left(
           \begin{array}{c}
        \l^\a        \\
         0      \\
   \end{array}
          \right),\quad
           \l_d=\left(
           \begin{array}{c}
        0       \\
         \l^\a      \\
   \end{array}
          \right)
 \eea
and similarly for other fields. Our conventions
 are summarized in Appendix A. The 32 by 32 gamma matrices are
\[
\G^m=\left(
           \begin{array}{cc}
        0  &  (\g^m)_{\a\b}        \\
        (\g^m)^{\a\b}  &   0      \\
   \end{array}
          \right)\;\;,\;\;
\]
The 16-component relation
 \bea
 (\l \g^m...)(... \g_m \l)=0
 \eea
translates into the 32-component relation
 \bea
 (\l_u \G^m...)(...\G_m \l_d)=0 \label{lggl}
 \eea
The following combination of $\l$-fields appears as a part of the
$[ds]$-integration measure
 \bea
 (\l\g_m)_{\k_1}(\l\g_n)_{\k_2}(\l\g_p)_{\k_3}\g^{mnp}_{\k_4\k_5}\;\;:
 \;\;\mbox{ anti-symmetric in $\k$'s}
 \label{anti-sym}
 \eea
 It is is totally antisymmetric in the $\k$-indices:
as we will see, it provides a very powerful freedom in various index
contractions. Throughout the computations we do not precisely record
the overall numerical coefficients and some of the irrelevant
non-numerical factors.\footnote{Not recording precise coefficients
is a common practice in the pure spinor literature.  In a high order
loop computation such as two- or (especially) three- loop, there are
many terms that need to be computed term by term. Many of these
terms actually vanish. Therefore it is important first to
efficiently identify the terms that lead to finite results. In
addition, we suspect that there is some subtlety in the AB
regulator. The upshot of section 4, can be put that one may obtain
the three-loop amplitude by deliberately not tracking the accurate
coefficients. This is because the AB regulator might produce a
vanishing amplitude due to pair-wise cancelations. Eventually one
should include not only the numerical coefficients but also the
functions of moduli parameters. For the two-loop, this was done in a
recent paper by Gomez and Mafra \cite{Gomez:2010ad}. }

 The
original one-loop computation was relatively simple. What makes it
even simpler is the introduction of new measures for, $[dr]$ and
$[ds]$ in \cite{Gomez:2009qd},
 \bea
 [dr]& \doteq & (\lb\g^m)^{\a_1}(\lb\g^n)^{\a_2}(\lb\g^p)^{\a_3}
          (\g_{mnp})^{\a_4\a_5}\e_{\a_1...\a_5\d_1...\d_{11}}
           \pa_r^{\d_1}\cdots \pa_r^{\d_{11}}
 \eea
 \bea
 [ds]&\doteq &\fr{1}{(\l\lb)^3}(\l\g^m)_{\a_1}(\l\g^n)_{\a_2}(\l\g^p)_{\a_3}
          (\g_{mnp})_{\a_4\a_5}\e^{\a_1...\a_5\d_1...\d_{11}}
           \pa^s_{\d_1}\cdots \pa^s_{\d_{11}} \nn\\
 \eea
These measures reduce to a great extent the amount of the algebra
involved in the two- and three- loops. For one- and two- loops, we
use the regulator given in \cite{Mafra:2009wq}
 \bea
\N_2= e^{-\lb\l-r\th-\wb w+sd} \label{2loopreg}
 \eea
 instead of the original regulator.
Unlike the latter, the former is not gauge invariant.\footnote{ It
was argued in \cite{Mafra:2009wq} that since $\N_2-1$ is BRST
trivial the amplitude would not be affected. For the three-loop,
however, things are not so clear, as we will discuss in section 4.}
For our purpose, it is only necessary to evaluate the $[ds^I]$- and
$[dd^I]$- integrations. The one-loop analysis goes as follows.
Saturating the 16 zero modes, the one-loop amplitude contains
 \bea
 {\cal K}&=&\int [d\l][d\lb][dr][d\th][dw][d\wb][ds][dd]
 e^{-\lb\l-r\th-\wb w+sd}
 \nn\\
&& \fr{(\lb\g^{mnp}\,r)(d\g_{mnp}d)}
    {192(\lb\l)^2}[\l^{\a_1} A_{\a_1}^\1]
 [d_{\a_2}W^{\a_2}_\2]
 [d_{\a_3}W^{\a_3}_\3][ d_{\a_4}W^{\a_4}_\4]
 \label{K}
 \eea
Carrying out $[ds]$ and $[dd]$ integration leads to (up to an
overall numerical factor)
 \bea
{\cal K}& \doteq &\int [d\l][d\lb][d\th][dw][d\wb]
 e^{-\lb\l-\wb w}(\lb\g^r)^{\a_1}(\lb\g^s)^{\a_2}(\lb\g^t)^{\a_3}
          (\g_{rst})^{\a_4\a_5}\e_{\a_1...\a_5\d_1...\d_{11}}
          \th^{\d_1}...\th^{\d_{11}}
 \nn\\
&& \fr{(\lb\g^{mnp}\,D)}
    {(\lb\l)^5}[\l A_\1]
 [\l \g^m W_\2]
 [\l \g^n W_\3][\l \g^p W_\4]
 \eea
which can be further evaluated to get the momentum space expression,
often called $K_0$, in the literature.

\subsection{two- and three- loop results}
The two- and three- loop computations follow similar, although more
complex, steps. After saturating the $d$-zero modes, one gets two-
and three- loop analogues of \rf{K}. While the one-loop expression,
\rf{K}, has the factor of $(d\g_{mnp}d)$, the corresponding two- and
three- loop expressions contain $(d\g_{mnp}d)^3$ and
$(d\g_{mnp}d)^5$ (or $(d\g_{mnp}d)^6$) respectively in a schematic
notation. The zero mode parts of these factors are
 \bea
 (d\g_{mnp}\,d)\doteq && (d_0\g_{mnp}\,d_0)\oplus(d_0'\g_{mnp}\,d_0')
                        \oplus(d_0\g_{mnp}\,d_0') \label{2loop_dd_quoted}
 \eea
for the two-loop and
 \bea
 (d\g_{mnp}\,d)\doteq && (d_0\g_{mnp}\,d_0)\oplus(d_0'\g_{mnp}\,d_0')
                       \oplus(d_0''\g_{mnp}\,d_0'')\nn\\
                     && \oplus(d_0\g_{mnp}\,d_0')\oplus(d_0\g_{mnp}\,d_0'')\oplus
                      (d_0'\g_{mnp}\,d_0'') \label{3loop_dd_quoted}
 \eea
for the three-loop. At some point of the evaluation of a given
amplitude, the zero-mode integrations need to be dealt with.
Obviously the higher powers of the factor $(d\g_{mnp}d)$ and the
larger numbers of the zero-modes are two of the causes that make a
higher loop computation more complex than a tree- or one- loop
amplitude.

One simplifying feature is that many of the $d$-zero structures that
result from expanding factors of $(d\g_{mnp}\,d)$ actually vanish.
This and some other aspects of the computation can be illustrated
with the two-loop case. The zero mode part of the two-loop factor,
$(d\g_{mnp}\,d)^3$, yields five different structures up to various
interchanges of $d_0$ and $d_0'$. The two-loop amplitude comes from
the following structure
 \bea
 &&(d_0\g_{m_1n_1p_1}\,d_0')(d_0\g_{m_2n_2p_2}\,d_0')(d_0\g_{m_3n_3p_3}\,d_0')\nn\\
 \label{d6_nonzero_q}
 \eea
which is shown to yield
 \bea
 (r\l) (\l \g^r W_\1)(\l \g^q W_\3)
 (r  W_\4) (r \g^q \g^{r} W_\2)  \label{tlfinal_q}
 \eea
in the next section. (Upon further evaluation, this yields the know
two-loop result, (\ref{2loopresult}).) Now we make the following
important observations that are believed to remain true for any
$d_0$-structure of any loop order. It is certainly true for the two-
and all of the three- loop terms that we analyze in the next two
sections. As mentioned above, the other four structures lead to
vanishing results. The fact that they vanish is not so obvious in
the beginning but becomes recognizable only after applying a serious
of identities. The method presented in section 3.1 corresponds to
keeping track of the overall numerical coefficient for a {\em given}
$d_0$-structure. There is more than one way to evaluate a given
$d_0$-structure. One may follow a more brute force type calculation
where typically the expression under consideration breaks into many
pieces. One can then simplify each contribution. The pattern that is
observed is that they either yield zero or exactly the same
expression as \rf{tlfinal_q}. What it means is that all of the five
structures yield an expression that is proportional to
\rf{tlfinal_q} in a generic method of computation. It is only one of
the structures given in \rf{d6_nonzero_q} that yields a
non-vanishing overall coefficient. We discuss this with an explicit
example in section 4. (See footnote 11 for instance.) The structure
of \rf{d6_nonzero_q} is ``tenacious": it is not so sensitive to the
$d_0$-structures. Another important feature of the computations is
that, typically, at the start of the simplification, the expression
has more indices than those of (53) where there are two indices,
r,q. All of the terms reduce to (53) or zero in the two-loop,
regardless of how many indices they had initially. This is what
happened in all examples analyzed in the three-loop case as well.
They all reduce to (103) or zero.

For the three-loop, there are additional complications that arise
from the regulator issue. Unlike the two-loop case, no explicit
gauge invariant regulator is known, at least not in a simple form.
We will discuss the complications that are due to this lack of
knowledge, but first will present the upshot of the analysis in the
section 4. Based on the aforementioned tenaciousness of the
amplitude structures, the following expression is proposed as an
anticipated three-loop amplitude,
 \bea
  (\l r)
 (\l \g^{t_1}\g^{t_2}r)[r W][\l \g^{t_3}W]
 [r \g^{t_1}\g^{m'} W][r \g^{t_2}\g^{t_3}W]
 \label{3loopresultq}
  \eea
  The expression is a
 three-loop analogue of (\ref{tlfinal_q}). It should be relatively
 straightforward to put it in the form that
 corresponds to (\ref{2loopresult}) and eventually further down to a full
 momentum space expression. For the remainder of this
 section, we outline the reasoning in section 4 that led to
 (\ref{3loopresultq}).

Setting apart the issue of gauge invariance, Berkovits and Aisaka
have proposed a relatively simple three-loop regulator
\cite{Aisaka:2009yp},
 \bea
 \N_3=&&\exp[-\sum_I^3 (\bar{w}_I^\a w_{\a,I}+s_I^\a d_{\a,I})]
      \exp(f^\a w_\a+g^\a d_\a+\bar{f}_\a \bar{w}^\a+\bar{g}_\a s^\a)
 \label{thlr_q}
 \eea
 elaborating on the work of \cite{Berkovits:2006vi}.
 The three-loop amplitude, ${\cal A}_{3-loop}$, given in
(\ref{loopprescription}) contains six insertions of the ghost field,
$b$. The reason that the two-loop regulator, (\ref{2loopreg}), is
not sufficient for the three-loop computation can be seen through a
``power counting" in $\fr{1}{\l\lb}$. The six $b$-insertions produce
several terms that would make the integration diverge. (This implies
$0\cdot \infty$ ambiguities because integrations over other fields
give a vanishing result. Nevertheless it is useful to evaluate the
three-loop with (\ref{2loopreg}) inserted. The full justification
will be given in section 4.) The \rf{thlr_q} regulates the
divergence by shifting $\l\lb$ by a constant. With use of
(\ref{thlr_q}), it is believed that the result of evaluation of the
${\cal A}_{3-loop}$ will be independent of the orders in which
various integrations are performed. By performing the $ds$- and
$dd$- integrations, it is easy to see that all the higher order
$\fr{1}{\l\lb}$-terms that have rendered the two-loop regulator
invalid do not actually contribute\footnote{It would be more natural
if the ${\cal A}_{3-loop}$ received some contributions from those
higher order terms. We will discuss this issue further in the
conclusion.}: they simply vanish due to the $d_0$-mode saturation
condition. This is curious since it will mean that the two-loop
regulator, (\ref{2loopreg}), will be equally good. In other words,
the results of an evaluation of the ${\cal A}_{3-loop}$ using
(\ref{2loopreg}) and (\ref{thlr_q}) are likely\footnote{There is a
subtlety. It will be discussed in section 4. } to be the same. The
result of the evaluation with the two-loop regulator turns out to
vanish as will be shown in section 4.1. This is an indication that
the three-loop regulator \rf{thlr_q} could lead to an entirely
vanishing result. A few other indications will be discussed in the
section 4.

It is worth recalling what happened in the two-loop case. In the
two-loop, even the structures that give vanishing results produce
the two-loop amplitude, (\ref{tlfinal_q}). It is just that the
overall coefficients vanish. Therefore the question is whether there
is a tenacious structure - analogous to \rf{tlfinal_q}- in the
three-loop as well. The answer is yes. In section 4, we take several
$d_0$-structures out of $(d \g_{mnp} d)^5$ (also out of $(d \g_{mnp}
d)^6$) and confirm that they all lead to zero or
(\ref{3loopresultq}). To see (\ref{3loopresultq}) produced, one
should follow a brute-force type analysis as in the two-loop
analysis; namely the analysis in section 3.2. The status of the
three-loop is different from that of the two-loop. First of all,
there is a possibility that the three-loop regulator, \rf{thlr_q},
is incomplete as pointed out above. Secondly, in general there are
far more terms to be evaluated. For one thing, there are more
$d_0$-structure simply because there are more $d_0$-modes as can be
seen from (\ref{3loop_dd_quoted}). The task of checking whether
\rf{thlr_q} indeed leads to a vanishing amplitude will require a
huge amount of algebra, but so does \rf{thlr_q}. Not only $(d
\g_{mnp} d)^5$ generates many sub-cases to be analyzed. Fortunately,
the tenaciousness that was observed in the two-loop analysis seems
to persist in the three-loop as well. From accumulated experience,
we believe that a certain pattern exists in the amplitude
computation. (It may be viewed as a feature of the pure spinor
formulation.) The pattern is that the amplitude structures such as
(\ref{tlfinal_q}) and (\ref{3loopresultq}) reside in many {\em
individual contributions even in their most broken down forms as
sub-cases}. It is {\em not} that several different $d_0$-structures
must be combined to get (\ref{tlfinal_q}) or (\ref{3loopresultq}).
If true, it means that the role of a proper (or a gauge invariant)
regulator is to make the overall coefficient non-vanishing. When
expanded, the regulator, \rf{thlr_q}, generates many cases to be
analyzed. In section 4, we take a few random cases and carry out the
analysis. Although we do not consider all the cases, those
considered cover many other cases partially because of the powerful
antisymmetry mentioned in (\ref{anti-sym}).

\section{Direct two-loop computation}

In pure spinor literature, the two-loop amplitude was first obtained
by a symmetry argument. Here we reproduce the result by a direct
computation. The main purpose of the direct computation is to set
the stage for a similar (but more complex) analysis of the
three-loop in the next section. In the two-loop case, there are 32
$d$-zero modes to saturate. The two-loop regulator, $\N_2$, provides
22 of them. The remaining $d$-zero modes should come from the
$b$-ghosts and the vertex operators: the two-loop amplitude is
 \bea
&& \int [d\l][d\lb][dr] [d^{16}\th] \prod_{I=1}^2
[dw^I] [d\bar w^I] [ds^I] [d^{16} d^I] \nn\\
 &&e^{-\lb\l-r\th-\wb{}w+sd}
 \nn\\
&& \fr{(\lb\g^{m_1n_1p_1}\,r)(d\g_{m_1n_1p_1}\,d)}
    {192(\lb\l)^2}\fr{(\lb\g^{m_2n_2p_2}\,r)(d\g_{m_2n_2p_2}\,d)}
    {192(\lb\l)^2}\fr{(\lb\g^{m_3n_3p_3}\,r)(d\g_{m_3n_3p_3}\,d)}
    {192(\lb\l)^2}\nn\\
 &&[d_{\a_1}W^{\a_1}_\1][d_{\a_2}W^{\a_2}_\2]
 [d_{\a_3}W^{\a_3}_\3][ d_{\a_4}W^{\a_4}_\4]
 \eea
where
 \bea
 (d\g_{m_in_ip_i}\,d)\sim \sum_{I,J}(d^I\g_{m_in_ip_i}\,d^J)
 \eea
with $d^I (I=1,2)$ denoting the zero modes. (We often write $d^I$
instead of $d_0^I$ for the simplicity of notation.) The factor,
$[d_{\a_1}W^{\a_1}_\1][d_{\a_2}W^{\a_2}_\2]
 [d_{\a_3}W^{\a_3}_\3][ d_{\a_4}W^{\a_4}_\4]$, will give the sum of
 terms with different integrations on the moduli parameters. As for the
 kinematic factor, one can choose appropriate combinations in order
 to have overall
 $16\;d^{I=1}$'s and $16\;d^{I=2}$'s respectively. After the
$[ds^I]$-integration, one gets
 \bea
 &&\fr{1}{(\lb \l)^3}\left[\fr{}{}\right. (\l \g^r)_{\a_1}(\l \g^s)_{\a_2}(\l \g^q)_{\a_3}
   (\g_{rsq})_{\a_4\a_5}\;\e^{\a_1...\a_5\r_1...\r_{11}}
   d_{\r_1}...d_{\r_{11}}\nn\\
 && \quad\quad \quad(\l \g^{r'})_{\a_1'}(\l \g^{s'})_{\a_2'}(\l \g^{q'})_{\a_3'}
   (\g_{r's'q'})_{\a_4'\a_5'}\;\e^{\a_1'...\a_5'\r_1'...\r_{11}'}
   d'_{\r_1'}...d'_{\r_{11}'}\left.\fr{}{}\right] \nn\\
 &&
 (d\g_{m_1n_1p_1}\,d)(d\g_{m_2n_2p_2}\,d)(d\g_{m_3n_3p_3}\,d)
 (\lb\g^{m_1n_1p_1}\,r)(\lb\g^{m_2n_2p_2}\,r)(\lb\g^{m_3n_3p_3}\,r)\nn\\
 &&\quad\quad\quad\quad\quad\quad\quad [d_{\a_1}W^{\a_1}_\1][d_{\a_2}W^{\a_2}_\2]
 [d_{\a_3}W^{\a_3}_\3][ d_{\a_4}W^{\a_4}_\4]\label{afterds}
 \eea
where in the abuse of the notation, we have defined
 \bea
 d^{I=1}&\equiv & d \nn\\
 d^{I=2}& \equiv & d'
 \eea
 To saturate the
$[dd^I]$-integral, 16 $d$'s and 16 $d'$'s must be present. Expansion
of
 \bea
 (d\g_{m_1n_1p_1}\,d)(d\g_{m_2n_2p_2}\,d)(d\g_{m_3n_3p_3}\,d)
 \eea
 in terms of the zero modes yields several terms. Each term comes
with an overall coefficient function of moduli parameters. In all of
the subsequent discussions, we omit those factors. There are five
different types of terms\footnote{There also terms that can be
obtained by interchanging the roles of $d$ and $d'$. At the end,
they will only change the overall numerical coefficient of the
computation of each type. }: the following types of terms
\bea
 &&(d\g_{m_1n_1p_1}\,d)(d\g_{m_2n_2p_2}\,d)(d'\g_{m_3n_3p_3}\,d')\nn\\
 &&(d\g_{m_1n_1p_1}\,d)(d\g_{m_2n_2p_2}\,d)(d\g_{m_3n_3p_3}\,d')\nn\\
 &&(d\g_{m_1n_1p_1}\,d)(d'\g_{m_2n_2p_2}\,d')(d\g_{m_3n_3p_3}\,d')\nn\\
 &&(d\g_{m_1n_1p_1}\,d')(d\g_{m_2n_2p_2}\,d')(d\g_{m_3n_3p_3}\,d)
 \label{dcomb}
 \eea
lead to vanishing results as we show in the section below. The fifth
type of term
 \bea
 &&(d\g_{m_1n_1p_1}\,d')(d\g_{m_2n_2p_2}\,d')(d\g_{m_3n_3p_3}\,d')\nn\\
 \label{d6_nonzero}
 \eea
produces the expected two-loop amplitude. We will work out that part
of the computation in the section that follows the next one.

\subsection{terms that vanish: analysis of eq.(\ref{dcomb}) }

We illustrate the vanishing of the terms in (\ref{dcomb}) with two
examples, the first term and the third term. While doing so, we
derive two identities that will be heavily used in the three-loop
analysis as well. The computation of the term that involves
  \bea
 (d\G_{m_1n_1p_1}\,d)(d\G_{m_2n_2p_2}\,d)(d'\G_{m_3n_3p_3}\,d')
 \eea
 goes as follows. Integrating over the measure $[ds]$, one gets
 \bea
 &&\fr{1}{(\lb \l)^3}\left[\fr{}{}\right. (\l \g^r)_{\a_1}(\l \g^s)_{\a_2}(\l \g^q)_{\a_3}
   (\g_{rsq})_{\a_4\a_5}\;\e^{\a_1...\a_5\r_1...\r_{11}}
   d_{\r_1}...d_{\r_{11}}\nn\\
 && \quad\quad \quad(\l \g^{r'})_{\a_1'}(\l \g^{s'})_{\a_2'}(\l \g^{q'})_{\a_3'}
   (\g_{r's'q'})_{\a_4'\a_5'}\;\e^{\a_1'...\a_5'\r_1'...\r_{11}'}
   d'_{\r_1'}...d'_{\r_{11}'}\left.\fr{}{}\right] \nn\\
 &&
 (d\g_{m_1n_1p_1}\,d)(d\g_{m_2n_2p_2}\,d)(d'\g_{m_3n_3p_3}\,d')
 (\lb\g^{m_1n_1p_1}\,r)(\lb\g^{m_2n_2p_2}\,r)(\lb\g^{m_3n_3p_3}\,r)\nn\\
 &&\quad\quad\quad\quad\quad\quad\quad [d_{\a_1}W^{\a_1}_\1][d_{\a_2}W^{\a_2}_\2]
 [d_{\a_3}W^{\a_3}_\3][ d_{\a_4}W^{\a_4}_\4]
 \label{rsq}
 \eea
 Performing  the $[dd]$-integration and utilizing the freedom
 mentioned in (\ref{anti-sym}), the relevant part of the above equation
  becomes
 \bea
 && (\l \g^r \g_{m_1n_1p_1} \g^s \l)(\l \g^q W^\1)
 \;\Tr(\g_{m_2n_2p_2}\g_{rsq})\nn\\
 && (\l \g^{r'} \g_{m_3n_3p_3} \g^{s'} \l)
     (W^\2\g_{r's'q'} W^\3)(\l \g^{q'}W^\4)\nn\\
 &&(\lb\g^{m_1n_1p_1}\,r)(\lb\g^{m_2n_2p_2}\,r)(\lb\g^{m_3n_3p_3}\,r)
 \eea
 In the 32-component notation, it takes the form of
 \bea
 && (\l_u \G^r \G_{m_1n_1p_1} \G^s \l_d)(\l_u \G^q W_d^\1)\nn\\
 && (\l_u \G^{r'} \G_{m_3n_3p_3} \G^{s'} \l_d)
     (W_u^\2\G_{r's'q'} W_d^\3)(\l_u \G^{q'}W_d^\4)\nn\\
 &&(\lb_d\G^{m_1n_1p_1}\,r_u)(\lb_d\G^{rsq}\,r_u)(\lb_d\G^{m_3n_3p_3}\,r_u)
 \label{dddddpdp}
 \eea
Using one of the 16-component Fierz identities, \rf{gamma3},
 \bea
 \g^m_{\a\b}\g^m_{\d\s}=-\fr12 \g^m_{\a\d}\g^m_{\b\s}
   +\fr1{24}\g^{mnp}_{\a\d}\g^{mnp}_{\s\b} \label{gamma3q}
 \eea
one can easily obtain the 16-component version of
 \bea
  (\lb_d\G^{mnp}\,r_u)(\l_u \G^r \G_{mnp} \G^s\l_d)
 &=& -48 (\lb_d  \G^s \G^r\l_d)(\l_u
 r_u)+48(\lb\l)(\l_u \G^r \G^s r_u) \label{fueproof2}
  \eea
 Both of these terms yield vanishing expressions. The first term
 with another factor in \rf{dddddpdp},
 $(\lb_d\G^{rsq}r_u)$, leads to a vanishing expression,
 \bea
 (\lb_d  \G^s \G^r\l_d)(\lb_d\G^{rsq}r_u)=0
 \eea
due to one of the pure spinor constraints.
 The second term of \rf{fueproof2} gives, up to an overall numerical
 factor,
 \bea
 && (\lb\l)^2 (\l_u \G^r  \G^s r_u) (\l_u \G^{r'}  \G^{s'} r_u)
 (\l_u \G^q W_d^\1)\nn\\
 &&
     (W_u^\2\G_{r's'q'} W_d^\3)(\l_u \G^{q'}W_d^\4)(\lb_d\G^{rsq}\,r_u)
     \label{vanishingexp}
 \eea
 It contains the factors $(\lb_d\G^{rsq}\,r_u)(\l_u \G^r  \G^s
 r_u)$. By Fierzing, it is not difficult to see that it vanishes:
 \bea
 (\lb_d\G^{rsq}\,r_u)(\l_u \G^r  \G^s r_u)=0 \label{keyid1}
 \eea
There are a few different ways to prove it. The proof that we adopt
shows the interplay between the 16- and 32- component notations. The
anti-symmetry of the first factor allows one to write the second
factor as $(\l_u \G^{rs} r_u)$. Because of the constraint,
$(\lb_d\G^{m}\,r_u)=0$, the left hand side can be rewritten as
 \bea
 &\sim& (\lb_d\G^q \G^{rs}\,r_u)(\l_u \G^{rs}   r_u)\nn\\
 &\sim& (\lb \g^q \g^{rs}\,r)(\l \g^{rs}   r)
  \label{keyid1prf}
 \eea
The 16-dimensional gamma matrix identity \cite{Berkovits:2006vi},
 \bea
 (\g^{mn})_\a{}^{\g}(\g_{mn})_\b{}^{\s}=-8\d_\a{}^{\s}\d_\b{}^{\g}
 -2\d_\a{}^{\g}\d_\b{}^{\s}+4 (\g^{p})_{\a\b}(\g_{p})^{\g\s}
 \eea
can now be used in the second equation of \rf{keyid1prf} to complete
the proof. To see the third term of \rf{dcomb}
\bea
 &&(d\g_{m_1n_1p_1}\,d)(d'\g_{m_2n_2p_2}\,d')(d\g_{m_3n_3p_3}\,d')
 \label{3rddcomb}
 \eea
vanish, let us use the identity \rf{gamma3} (quoted in \rf{gamma3q})
in
\bea
 &&(\g_{m_1n_1p_1})^{\k_1\k_2}(\g_{m_2n_2p_2})^{\k_1'\k_2'}
 (\g_{m_3n_3p_3})^{\k_3\k_3'}\nn\\
 &&(\lb\g^{m_1n_1p_1}\,r)(\lb\g^{m_2n_2p_2}\,r)(\lb\g^{m_3n_3p_3}\,r)
 \eea
 where the $d$-factors have been suppressed.
 Taking the pure spinor constraint, $(\lb \g^p r)=0$, the above
 equation simplifies to
 \bea
 (r \g^{u_1})^{\k_1} (\lb\g_{u_1})^{\k_2}
 (r \g^{u_2})^{\k_1'}(\lb\g_{u_2})^{\k_2'}
 (r \g^{u_3})^{\k_3} (\lb\g_{u_3})^{\k_3'}
 \eea
 After performing $[ds]$- and $[dd]$- integrations, one can
choose (again thanks to the freedom stated in (\ref{anti-sym})) the
contractions with $\l$-factors to get
 \bea
 &&(r\g^{rsq}\lb)(r\g^{r's'q'}\lb)(r\g^{u_3}\g^r \l)
  (\lb \g_{u_3}\g^{r'}\l) \nn\\
 &&\sim (\lb\l)(r \g^{rsq}\lb) (r \g^{r's'q'}\lb)
     (r\g^{r'}\g^r \l)
 \eea
Since the factor $(d\g_{m_3n_3p_3}\,d')$ in (\ref{3rddcomb}) can be
written as $(d'\g_{m_3n_3p_3}\,d)$, the factor, $(r\g^{r'}\g^r \l)$,
can be replaced by $\eta^{r'r}(r\l)$. Once used in the equation
above, the resulting expression vanishes due to the constraint,
$(...\g^r\lb)(...\g_r \lb)$=0.

\subsection{the two-loop amplitude from eq.\rf{d6_nonzero}}

The two-loop amplitude is produced by the term in \rf{d6_nonzero}
 \bea
 (d\g_{m_1n_1p_1}\,d')(d\g_{m_2n_2p_2}\,d')(d\g_{m_3n_3p_3}\,d')
 \eea
 After carrying out the $d$-integration, one can choose the index
 contractions appropriately so as to maximize the usage of
 \rf{fueproof2}. There are several different type of terms depending
 on the frequency of appearance of either the first or the second
 term of (\ref{fueproof2}). The first term cannot appear more than
 once since otherwise the resulting expression would contain
 \bea
 (\l r)(\l r)=0
 \eea
 due to the (anti-)commutativity of
 $\l$ and $r$.
Therefore there are two types of terms
 \bea
 &&(\lb\l)^2 (\l r) (\lb \g^{r'}\g^r \l)(\l \g^s\g^{s'} r)
 (\l \g^q\g^{q'} r)
 [W^\1\g_r\g_s\g_q W^\2][W^\3\g_{r'}\g_{s'}\g_{q'}W^\4],\nn\\
 &&(\lb\l)^3 (\l \g^r\g^{r'} r)(\l \g^s\g^{s'} r)
 (\l \g^q\g^{q'} r)
  [W^\1\g_r\g_s\g_q W^\2][W^\3\g_{r'}\g_{s'}\g_{q'}W^\4]
 \label{primitive_startingpoint}
 \eea
 Up to an overall numerical factor they give the same
 contributions: We illustrate the computation with the second term.
 To apply \rf{etagg}, we rewrite the second term,
 \bea
 &&(\lb\l)^3\; \l^{\a_1}\; (\g^{r'})^{\a_2\a_3}\; r_{\a_3}\;
   \l^{\b_1}\; (\g^s)_{\b_1\b_2}\; r_{\b_3}\;
 \l^{\g_1} (\g^{q'})^{\g_2\g_3} r_{\g_3}\nn\\
 && {W_\1}^{\r_1}(\g_s)^{\r_2\r_3} {W_\2}^{\r_4}
  {W_\3}^{\s_1}(\g_{r'})_{\s_1\s_2}(\g_{q'})_{\s_3\s_4}W_\4^{\s_4}\nn\\
 &&(\g^r)_{\a_1\a_2}(\g_r)_{\r_1\r_2}(\g^q)_{\g_1\g_2}
   (\g_q)_{\r_3\r_4}(\g^{s'})^{\b_2\b_3}(\g_{s'})^{\s_2\s_3}
 \eea
Some of the terms that result by using \rf{etagg} in the third line
vanish. The surviving terms can be put into
 \bea
  (r\l)(\l \g^r W_\1)(\l \g^q W_\3)
 (r  W_\4) (r \g^q \g^{r} W_\2)  \label{tlfinal}
 \eea
Up to an irrelevant overall numerical factor, it can be put into
 \bea
  &&(\l \g^r )_{\b_1}(\l \g^q )_{\b_3}
 ( \g^q \g^{r} )^{\a_2}{}_{\b_2}\; \l^{\a_1} \nn\\
 && D_{\a_2}D_{\a_1}D_{\b_4}\;
   \left[W_\1^{\b_1}W_\2^{\b_2}W_\3^{\b_3}W_\4^{\b_4}\right]
 \eea
There are three types of terms depending how the covariant
derivatives are distributed: the first type of terms is such that
all $D$'s act on different $W's$. (Recall that due to the field
equation, one has $D_{\b_4}W_\4^{\b^4}=0$.) These are the terms that
produce the non-zero result as we will discuss shortly. The other
two types of terms are the ones with all three $D$'s acting on the
same $W$ and the ones with two $D$'s acting on the same $W$ and the
third $D$ acting on another $W$. They all vanish due to either
Bianchi identity or the presence of a vanishing factor $(\l \g^{mnp}
\l)$.

Let us consider each type of the term. The first is the type of the
terms where all three $D$'s act on the same $W$. The terms that
contain $D_{\a_2}D_{\a_1}D_{\b_4} W_\1^{\b_1}$ lead to
 \bea
 \doteq k_{l_1}^\1 \F_{v_1v_2}^\1(\l \g^{v_1} W_\2)
 (\l \g^{v_2} W_\3)(\l \g^{l_1} W_\4) \label{DDDW1}
 \eea
 The indices, $(l_1,v_1,v_2)$, will be anti-symmetrized once the permutations
 in $(\1,\2,\3,\4)$ are taken into account, so the equation
 above vanishes due to Bianchi identity. The term containing
$D_{\a_2}D_{\a_1}D_{\b_4} W_\2^{\b_2}$ directly leads to
$k_{[l_1}^\2 \F_{v_1v_2]}^\2$, therefore vanishes. The terms
containing $D_{\a_2}D_{\a_1}D_{\b_4} W_\3^{\b_3}$ are given by
$(1\leftrightarrow 3)$ of \rf{DDDW1}. The terms with two $D$'s
acting on the same $W$ and the third $D$ acting on another $W$
produces a similar result. For example the term with $
(D_{\a_1}D_{\b_4} W_\1^{\b_1}) (D_{\a_2} W_\2^{\b_2}) W_\3^{\b_3}
W_\4^{\b_4}$ yields
 \bea
 \doteq (\l \g^m W^\1)(\l \g^n W^\3)(\l \g^p W^\4)
   k^\1_p\F^\2_{mn}
 \eea
Consider another example, $ (D_{\a_1}D_{\b_4} W_\1^{\b_1})
W_\2^{\b_2} (D_{\a_2}W_\3^{\b_3}) W_\4^{\b_4}$. It yields
 \bea
 \doteq (\l \g^m W^\4)(\l \g^u W^\2)(\l \g^v \pa_m W^\1)
  \F^\3_{uv}
 \eea
There are altogether 27 terms of this type.
After some algebra one can show
 \bea
 && -D_{\a_1}D_{\b_4} W_\1^{\b_1} D_{\a_2} W_\2^{\b_2}
      W_\3^{\b_3} W_\4^{\b_4}+D_{\a_1}D_{\b_4} W_\1^{\b_1} W_\2^{\b_2}
       D_{\a_2}W_\3^{\b_3} W_\4^{\b_4} \nn\\
    &&\hspace{2.3in} -D_{\a_1}D_{\b_4} W_\1^{\b_1}  W_\2^{\b_2}
      W_\3^{\b_3} D_{\a_2}W_\4^{\b_4}\nn\\
 &&+D_{\a_2}D_{\b_4} W_\1^{\b_1} D_{\a_1} W_\2^{\b_2}
      W_\3^{\b_3} W_\4^{\b_4}+D_{\b_4} W_\1^{\b_1} D_{\a_2}D_{\a_1} W_\2^{\b_2}
      W_\3^{\b_3} W_\4^{\b_4}\nn\\
 &&-D_{\a_2}D_{\b_4} W_\1^{\b_1} W_\2^{\b_2}
      D_{\a_1} W_\3^{\b_3} W_\4^{\b_4}+D_{\b_4} W_\1^{\b_1} W_\2^{\b_2}
      D_{\a_2}D_{\a_1} W_\3^{\b_3} W_\4^{\b_4}\nn\\
 &&+D_{\a_2}D_{\b_4} W_\1^{\b_1} W_\2^{\b_2}
       W_\3^{\b_3} D_{\a_1}W_\4^{\b_4}+D_{\b_4} W_\1^{\b_1} W_\2^{\b_2}
       W_\3^{\b_3} D_{\a_2}D_{\a_1}W_\4^{\b_4}\nn\\
 && -D_{\a_2}D_{\a_1} W_\1^{\b_1}D_{\b_4} W_\2^{\b_2}
       W_\3^{\b_3} W_\4^{\b_4}-D_{\a_1} W_\1^{\b_1}D_{\a_2}D_{\b_4} W_\2^{\b_2}
       W_\3^{\b_3} W_\4^{\b_4}\nn\\
  &&+D_{\a_2} W_\1^{\b_1}D_{\a_1}D_{\b_4} W_\2^{\b_2}
       W_\3^{\b_3} W_\4^{\b_4}+ W_\1^{\b_1}D_{\a_1}D_{\b_4}D_{\b_4} W_\2^{\b_2}
       D_{\a_2}W_\3^{\b_3} W_\4^{\b_4}\nn\\
  &&\hspace{2.3in}     -W_\1^{\b_1}D_{\a_1}D_{\b_4} W_\2^{\b_2}
       W_\3^{\b_3} D_{\a_2}W_\4^{\b_4}\nn\\
  &&-W_\1^{\b_1}D_{\a_2}D_{\b_4}W_\2^{\b_2}
       D_{\a_1}W_\3^{\b_3} W_\4^{\b_4}-W_\1^{\b_1}D_{\b_4} W_\2^{\b_2}
       D_{\a_2}D_{\a_1}W_\3^{\b_3} W_\4^{\b_4}\nn\\
  &&+W_\1^{\b_1}D_{\a_2}D_{\b_4} W_\2^{\b_2}
       W_\3^{\b_3} D_{\a_1}W_\4^{\b_4}-W_\1^{\b_1}D_{\b_4} W_\2^{\b_2}
       W_\3^{\b_3} D_{\a_2}D_{\a_1}W_\4^{\b_4}\nn\\
  &&+D_{\a_2}D_{\a_1}W_\1^{\b_1} W_\2^{\b_2}
       D_{\b_4}W_\3^{\b_3} W_\4^{\b_4}-D_{\a_1}W_\1^{\b_1} W_\2^{\b_2}
       D_{\a_2}D_{\b_4}W_\3^{\b_3} W_\4^{\b_4}\nn\\
  &&+W_\1^{\b_1}D_{\a_2}D_{\a_1} W_\2^{\b_2}
       D_{\b_4}W_\3^{\b_3} W_\4^{\b_4}+W_\1^{\b_1}D_{\a_1} W_\2^{\b_2}
       D_{\a_2}D_{\b_4}W_\3^{\b_3} W_\4^{\b_4}\nn\\
  &&+D_{\a_2}W_\1^{\b_1} W_\2^{\b_2}D_{\a_1}D_{\b_4}
        W_\3^{\b_3} W_\4^{\b_4}-W_\1^{\b_1} D_{\a_2}W_\2^{\b_2}D_{\a_1}D_{\b_4}
        W_\3^{\b_3} W_\4^{\b_4}\nn\\
   && \hspace{2.3in}-W_\1^{\b_1} W_\2^{\b_2}D_{\a_1}D_{\b_4}
        W_\3^{\b_3} D_{\a_2}W_\4^{\b_4}\nn\\
   &&+W_\1^{\b_1} W_\2^{\b_2}D_{\a_2}D_{\b_4}
        W_\3^{\b_3} D_{\a_1}W_\4^{\b_4}+W_\1^{\b_1} W_\2^{\b_2}D_{\b_4}
        W_\3^{\b_3} D_{\a_2}D_{\a_1}W_\4^{\b_4}\nn\\
  =&& (\l\g^mW_\1)(\l\g^nW_\3)(\l\g^pW_\4)
      \left[-6 k^\2_p \F^\2_{mn}
      -4k^\4_n\F^\2_{mp}+4k^\4_m\F^\2_{np}\right]\nn\\
  &&+ (\l\g^mW_\1)(\l\g^nW_\2)(\l\g^pW_\4)
      \left[-6 k^\3_p \F^\3_{mn}
      +4k^\4_m\F^\3_{np}-4k^\4_n\F^\3_{mp}\right]\nn\\
  &&+ (\l\g^mW_\1)(\l\g^nW_\2)(\l\g^pW_\3)
      \left[-2 k^\4_p \F^\4_{mn}+2 k^\4_n \F^\4_{mp}
      -2 k^\4_m \F^\4_{np}
      \right]\nn\\
  &&+(\l\g^mW_\2)(\l\g^nW_\3)(\l\g^pW_\4)
      \left[-6 k^\1_p \F^\1_{mn}-4k^\4_n\F^\1_{mp}+4k^\4_m\F^\1_{np}\right]
 \eea
The third term vanishes due to Bianchi identity, $k_{[m}\F_{np]}=0$.
The terms with the factors, $k^\1_p \F^\1_{mn},k^\2_p
\F^\2_{mn},k^\3_p \F^\3_{mn}$, vanish after taking the permutations
over (1,2,3,4) into account. The remaining terms are
 \bea
 && (\l\g^mW_\1)(\l\g^nW_\3)(\l\g^pW_\4)
      \left[
      -4k^\4_n\F^\2_{mp}+4k^\4_m\F^\2_{np}\right]\nn\\
  &&+ (\l\g^mW_\1)(\l\g^nW_\2)(\l\g^pW_\4)
      \left[
      4k^\4_m\F^\3_{np}-4k^\4_n\F^\3_{mp}\right]\nn\\
  &&+(\l\g^mW_\2)(\l\g^nW_\3)(\l\g^pW_\4)
      \left[-4k^\4_n\F^\1_{mp}+4k^\4_m\F^\1_{np}\right]
 \eea
The second term is obtained from the first term by
$(2\leftrightarrow 3)$ and the third term is obtained from the first
term by $(1\leftrightarrow 2)$. Therefore we may consider the first
term only. The first term is symmetric under $(1\leftrightarrow 3)$:
it is sufficient to consider
 \bea
 && (\l\g^mW_\1)(\l\g^nW_\3)(\l\g^pW_\4)
      k^\4_n\F^\2_{mp}
 \eea
 with its permutations in (1,2,3,4), altogether 24 terms. We
put them into four groups: the first group includes the terms with
$\F^\1_{mp}$, the second group $\F^\2_{mp}$, the third group
$\F^\3_{mp}$ and the fourth group $\F^\2_{mp}$. Each group has a
vanishing result. We illustrate this with the second group. The six
terms in the second group are
 \bea
 && (\l\g^mW_\1)(\l\g^nW_\3)(\l\g^pW_\4)\F^\2_{mp}k^\4_n
 +(\l\g^mW_\1)(\l\g^nW_\4)(\l\g^pW_\3)\F^\2_{mp}k^\3_n\nn\\
 &&+(\l\g^mW_\3)(\l\g^nW_\4)(\l\g^pW_\1)\F^\2_{mp}k^\1_n
 +(\l\g^mW_\3)(\l\g^nW_\1)(\l\g^pW_\4)\F^\2_{mp}k^\4_n\nn\\
 &&+(\l\g^mW_\4)(\l\g^nW_\1)(\l\g^pW_\3)\F^\2_{mp}k^\3_n
 +(\l\g^mW_\4)(\l\g^nW_\3)(\l\g^pW_\1)\F^\2_{mp}k^\1_n\nn\\
 \doteq &&
 (\l\g^mW_\1)(\l\g^nW_\3)(\l\g^pW_\4)k^\2_{[m}\F^\2_{np]}\nn\\
 =&&0
 \eea
where, in the second equality, the momentum conservation and$\!\ks
W=0$ have been used.
\\
Finally we illustrate with $ (D_{\b_4} W_\1^{\b_1}) (D_{\a_1}
W_\2^{\b_2})\, W_\3^{\b_3}\,(D_{\a_2} W_\4^{\b_4})$ the computations
that involve the terms with all three $D$'s acting on the different
$W$'s. Using the third eq. of \rf{fieldeqs}, the equation above
becomes
 \bea
 \doteq (\l \g^{m_3n_3}\g^{m_1n_1}\g^{m_2}\l)(\l \g^qW_\3 )
 \F_{m_1n_1}^\1\F_{m_2q}^\2\F_{m_3n_3}^\4
 \eea
After a series of manipulations using the identities given in
Appendix B and the Fierz rearrangement identities, it becomes the
desired expression
 \bea
 (\l \g^{\m_1...\m_5}\l)(\l \g^s W_\3)
 \F_{\m_1\m_2}^\1\F_{\m_3\m_4}^\4\F_{\m_5s}^\2
 \label{2loopresult}
 \eea

\section{Three-loop analysis}

The prescription for the three-loop amplitude (the main object of
the present work) is given in (\ref{loopprescription}) and quoted
here for convenience,
 \bea
  {\cal A}_{3-loop}&=&\int d\t_1d\cdots d\t_6 <{\cal N}_3(y)
 \prod_{s=1}^6\int dw_s \m(w_s)b(w_s)
 \int dz_1 U_1(z_1)\cdots  \int dz_N U_N(z_N)>\nn\\
 \eea
The two-loop regulator
 \bea
 \N_2= e^{-\lb\l-r\th-\wb w+sd} \label{tlr}
 \eea
does not properly carry out the $\l\lb\sim 0$ regulation in the case
of the three-loop. The six $b$-ghosts produce several potentially
divergent terms. Although most of these terms get removed by the
$d$-integration, the result of the computation depends on the order
of the integrations. Namely, if one performs the $\l$-integration
prior to the $d$-integration, one gets an ambiguous expression. One
possible resolution was discussed by Berkovits and Nekrasov
\cite{Berkovits:2006vi}. The idea is to shift $\l $ and $\lb$ so
that the singularity in $\fr{1}{\l\lb}$ can be avoided. The task
becomes more complicated in order to make the shifted spinors
satisfy the pure spinor constraints. Elaborating on the idea of
\cite{Berkovits:2006vi}, Aisaka and Berkovits proposed in
\cite{Aisaka:2009yp} a regulator
 \bea
 \N_3=&&\exp[-\sum_I^3 (\bar{w}_I^\a w_{\a,I}+s_I^\a d_{\a,I})]
      \exp(f^\a w_\a+g^\a d_\a+\bar{f}_\a \bar{w}^\a+\bar{g}_\a s^\a)
 \label{thlr}
 \eea
The regulator is not gauge invariant. It was introduced to serve a
specific purpose: to show how the regularization works in a
simplified setup. In the next section, we take the two-loop
regulator, \rf{tlr}, in the three-loop environment and carry out
relatively simple computations. The justification for using a
two-loop regulator in the three-loop computations will be given in
the beginning of that section. In the section that follows, we
examine the calculations that involve \rf{thlr}.

A few remarks are in order. Some of the higher power terms in
$\fr1{\l\lb}$ in the amplitude make the two-loop regulator invalid
based on a ``power counting". As we will see below, those terms do
not contribute to the computations with \rf{thlr} inserted.
Therefore, to the extent that they are genuine regulators it should
not matter which regulator to use. This being the case, there is a
problem since the computation with the two-loop regulator leads to a
vanishing result. Therefore, the complete analysis with \rf{thlr}
inserted could also lead to a vanishing result. Fortunately, this
undesirable possibility does not seem to prevent one from acquiring
the final forms\footnote{A form that is a three-loop analogue of
\rf{tlfinal} is meant.} of the terms in a given amplitude. This is
because of the way the calculations in the pure spinor formulation
work. What happens in the two-loop case may be useful for this
point. In the two-loop case, the $d$-structures given in \rf{dcomb}
yield vanishing results. What causes the vanishing in some cases is
pairwise cancelations among different terms. They produce the
correct form given in \rf{tlfinal} when evaluated in a brute-force
type calculation where no attention is paid to the overall numerical
coefficient.\footnote{A specific example may be helpful. In the
section 3.1, eq.\rf{vanishingexp} was shown to vanish due to the
identity given in (\ref{keyid1}). One could have used the identity
\rf{etagg}, instead, in the manner that is similar to that of the
section 3.2 where the two-loop amplitude, \rf{tlfinal}, is derived.
By doing so, one amounts to not tracking the overall numerical
coefficient. One can show after some algebra the resulting
expression is proportional to \rf{tlfinal}. } It is just that the
overall numerical coefficient vanishes. We believe that it may well
be a general feature of the pure spinor formulation.\footnote{ It is
a reasonable possibility that the role of a gauge invariant
regulator is merely to make the overall coefficient ``certainly
non-vanishing".} In section 4.2, we select a few representative
examples of the $d$-structures. We simplify them to the expressions
that are suitable for the further momentum space evaluation, a task
that we intend to take in the near future. Strictly speaking, the
three-loop amplitude proposed in (\ref{3loopresult}) is not a
rigorous result in the sense that not all the contributions are
analyzed.

\subsection{warm-up computation}

 The complete analysis using \rf{thlr} is tedious and
lengthy. It makes repeated use of a fixed set of techniques. In this
section, we display those techniques in a more controlled and less
diverse environment by using the two-loop regulator, \rf{tlr}, in
the three-loop calculation. There are two main reasons for this.
Firstly, the essential part of the three-loop calculation with
\rf{tlr} inserted will re-emerge in the three-loop calculation with
\rf{thlr} inserted. This will become clear in the next section. The
second reason is more theoretical. Inserted in the three-loop, the
two-loop regulator, \rf{tlr},
 does not regulate some of the terms that come in higher powers of
 $\fr{1}{\l\lb}$ whereas the regulator in \rf{thlr} does. Howevefr, it
 regulates them to zero as can be seen by considering the $d$-integration
 prior to the $\l$-integration. Therefore, it should not matter
 which regulator to use to carry out the integration of the terms of
 lower powers in $\fr{1}{\l\lb}$.\footnote{There is a subtlety.
 The logic assumes that \rf{tlr} and \rf{thlr} are genuine regulators.
 According to \cite{Hoogeveen:2007tu} they may
 originate from gauge fixing.}

 For
the three-loop, we will need 48 $d$-zero modes. Inspection of the
integration over the $d$-zero modes reveals that there are three
potentially non-zero contributions,\footnote{As pointed out by H.
Gomez, there could be contributions from
 \bea
  \left(\fr{(\lb\g^{mnp}\,r)(d\g_{mnp}\,d)}
    {192(\lb\l)^2}\right)^{6}
 (d_{\a_1}W_\1^{\a_1})(d_{\a_2}W_\2^{\a_2})
    (d_{\a_3}W_\3^{\a_3})( d_{\a_4}W_\4^{\a_4})
    \label{gomez}
 \eea
after one of $d$'s in $(d\g_{mnp}\,d)^6$ gets removed via OPE with
one the $W$'s in $(dW)^4$. The analysis of $(d\g_{mnp}\,d)^5$-case
in \rf{3loop} below can be applied to this case with only minimal
modifications.
 }
 \bea
 && \int [d\l][d\lb][dr] d^{16}\th \prod_{I=1}^3
[dw^I] [d\bar w^I] [ds^I] d^{16} d^I e^{-\lb\l-r\th-\wb w+sd}
 \nn\\
 &&\left[\fr{}{}\right.(d_{\a_1}W_\1^{\a_1})(d_{\a_2}W_\2^{\a_2})
 (d_{\a_3}W_\3^{\a_3})( d_{\a_4}W_\4^{\a_4})\nn\\
 &&\;\;+ \left(\fr{(\lb\g^{mnp}\,r)(d\g_{mnp}\,d)}
    {192(\lb\l)^2}\right)^{6}
      (\pa\th^{\a_1} A_{\a_1}+\Pi^{m_1} A_{m_1}
   +\fr12N^{m_1n_1}{\cal F}_{m_1n_1})\nn\\
  &&\quad\quad(d_{\a_2}W_\2^{\a_2})
 (d_{\a_3}W_\3^{\a_3})( d_{\a_4}W_\4^{\a_4})
 \left.\fr{}{}\right]\nn\\
 &&\;\;+ \left(\fr{(\lb\g^{mnp}\,r)(d\g_{mnp}\,d)}
    {192(\lb\l)^2}\right)^{5}(\lb\g^{m'}d)
    \left(\fr{\Pi^{m'}}{2(\lb\l)}
       -\fr{(r\g_{m'n'p'}r)N^{n'p'}}{16(\lb\l)^3}
            \right)\nn\\
     &&\quad\quad(d_{\a_1}W_\1^{\a_1})(d_{\a_2}W_\2^{\a_2})
 (d_{\a_3}W_\3^{\a_3})( d_{\a_4}W_\4^{\a_4})
 \label{3loop}
 \eea
The first term in the square bracket is the case where the regulator
${\cal N}_2$ should provide 44 $d$-zero modes. The six $b$-ghosts
must not provide any additional $d$-zero modes.
 The vertex operators provide four of them.
 \bea
 && \int [d\l][d\lb][dr] d^{16}\th \prod_{I=1}^3
[dw^I] [d\bar w^I] [ds^I] d^{16} d^I \nn\\
 &&e^{-\lb\l-r\th-\wb w+sd}
 \nn\\
 &&[d_{\a_1}W^{\a_1}(z_1)][d_{\a_2}W^{\a_2}(z_2)]
 [d_{\a_3}W^{\a_3}(z_3)][ d_{\a_4}W^{\a_4}(z_4)]
 \eea
 It is easy to see that this vanishes simply by carrying out the
 $[ds]$-integration. Some of the vanishing contributions here and below may be
 seen by a symmetry argument as well that is similar to the one
 in \cite{Berkovits:2006bk}. For
the second term in \rf{3loop}, the six $b$-ghosts should provide 12
remaining $d$'s and the vertex operators 3,
 \bea
 && \int [d\l][d\lb][dr] d^{16}\th \prod_{I=1}^g
[dw^I] [d\bar w^I] [ds^I] d^{16} d^I \nn\\
 &&e^{-\lb\l-r\th-\wb w+sd}
\left[\fr{(\lb\g^{mnp}\,r)(d\g_{mnp}\,d)}
    {192(\lb\l)^2}\right]^{6} \nn \\
 &&[\pa\th^{\a_1} A_{\a_1}+\Pi^{m_1} A_{m_1}
   +\fr12N^{m_1n_1}{\cal F}_{m_1n_1}][d_{\a_2}W^{\a_2}(z_2)]
 [d_{\a_3}W^{\a_3}(z_3)][ d_{\a_4}W^{\a_4}(z_4)]\nn\\
 \label{nonzero3}
 \eea
 Let us focus on the following factors,
 \bea
 \left[{(\lb\g^{mnp}\,r)(d\g_{mnp}\,d)}
   \right]^{6} \label{dd6}
 \eea
 Each factor of $(d\g_{mnp}\,d)$ consists of several combinations of
the zero modes
 \bea
 (d\g_{mnp}\,d)\doteq && (d_0\g_{mnp}\,d_0)\oplus(d_0'\g_{mnp}\,d_0')
                       \oplus(d_0''\g_{mnp}\,d_0'')\nn\\
                     && \oplus(d_0\g_{mnp}\,d_0')\oplus(d_0\g_{mnp}\,d_0'')\oplus
                      (d_0'\g_{mnp}\,d_0'') \label{dd}
 \eea
where the zero mode subscripts have been temporarily re-inserted.
Let us call the terms in the first line the diagonal terms and the
ones in the second line the cross terms,
 \bea
 &&(d_0^I\g_{mnp}\,d_0^I):\;\;\quad \mbox{a diogonal term} \nn\\
  &&(d_0^I\g_{mnp}\,d_0^J),\; I\neq J:\; \mbox{a cross term}
 \eea
Eq.\rf{dd6} produces several different types of terms depending
which factors in \rf{dd} are present. Before getting into specific
cases, a general structural analysis might be useful. Consider
\rf{3group} and break, into three groups, the expression that
results from the $d$-integration:
 \bea
\mbox{I}:  &&\left[\fr{}{}\right. (\l \g^r)_{\a_1}(\l
\g^s)_{\a_2}(\l \g^q)_{\a_3}
   (\g_{rsq})_{\a_4\a_5}\;
   \nn\\
 &&\;\;(\l \g^{r'})_{\a_ 1'}(\l \g^{s'})_{\a_2'}(\l \g^{q'})_{\a_3'}
   (\g_{r's'q'})_{\a_4'\a_5'}\;
   \nn\\
 &&\;\;(\l \g^{r''})_{\a_ 1''}(\l \g^{s''})_{\a_2''}(\l \g^{q''})_{\a_3''}
   (\g_{r''s''q''})_{\a_4''\a_5''}\;
   \left.\fr{}{}\right], \nn\\
II: &&
(\g_{m_1n_1p_1})^{\k_1\k_2}(\g_{m_2n_2p_2})^{\k_3\k_4}(\g_{m_3n_3p_3})^{\k_5\k_6}
    (\g_{m_4n_4p_4})^{\k_7\k_8}(\g_{m_5n_5p_5})^{\k_9\k_{10}}
    (\g_{m_6n_6p_6})^{\k_{11}\k_{12}}\nn\\
     && (\lb_{\s_1}(\g^{m_1n_1p_1})^{\s_1\s_2}\,r_{\s_2})
     (\lb_{\s_3}(\g^{m_2n_2p_2})^{\s_3\s_4}\,r_{\s_4})
     (\lb_{\s_5}(\g^{m_3n_3p_3})^{\s_5\s_6}\,r_{\s_6})\nn\\
      && (\lb_{\s_7}(\g^{m_4n_4p_4})^{\s_7\s_8}\,r_{\s_8})
     (\lb_{\s_9}(\g^{m_5n_5p_5})^{\s_9\s_{10}}\,r_{\s_{10}})
      (\lb_{\s_{11}}(\g^{m_6n_6p_6})^{\s_{11}\s_{12}}\,r_{\s_{12}}),
      \nn\\
 III: &&[W^{\b_2}(z_2)]
 [W^{\b_3}(z_3)][W^{\b_4}(z_4)]
 \eea
 We have suppressed factors of Kronecker deltas that dictate
 contractions of $(\k,\b)$-indices with $\a$-indices.
Appling the Fierz identity, \rf{gamma3}, and using one the pure
spinor constraints, $(\lb \g^\m r) =0$, the group (II) becomes
 \bea
  &&[(r \g^{u_1})^{\k_1} (\lb \g_{u_1})^{\k_2}][ (r \g^{u_2})^{\k_3}
   (\lb \g_{u_2})^{\k_4}][ (r \g^{u_3})^{\k_5}
   (\lb \g_{u_3})^{\k_6}]\nn\\
  && [(r \g^{u_4})^{\k_7} (\lb
   \g_{u_4})^{\k_8}][(r \g^{u_5})^{\k_9} (\lb
   \g_{u_5})^{\k_{10}}]
   [(r \g^{u_6})^{\k_{11}} (\lb
   \g_{u_6})^{\k_{12}}]\label{g3id}
 \eea
These factors get contracted with the factors in the group (I) in
various ways. The resulting expressions allow further simplification
by way of the pure spinor constraints and the gamma matrix
properties. We illustrate this with a few examples. When there are
two or more of
 the diagonal terms, $(d^I\g_{mnp}\,d^I)$ with the given $I$, the
 contribution trivially vanishes due to the index structure.
 The following type also vanishes
 \bea
 (d^I\g_{m_1n_1p_1}\,d^J)(d^I\g_{m_2n_2p_2}\,d^J)...
 \eea
For example, consider
$(d\g_{m_1n_1p_1}\,d')(d\g_{m_2n_2p_2}\,d')\cdots$. It implies that
(\ref{g3id}) contains
 \bea
 [(r \g^{u_1})^{\k_1} (\lb \g_{u_1})^{\k_1'}]
 [(r \g^{u_2})^{\k_2} (\lb \g_{u_2})^{\k_2'}]
 \eea
 The contractions with the group (I) factors can be chosen in such a way that
 the resulting expression contains
 \bea
 (r\g^{u_1}\g^{rsq}\g^{u_2}r) (\lb\g^{u_1}\g^{r's'q'}\g^{u_2}\lb)
 \eea
It vanishes because the first factor is symmetric under
$u_1\leftrightarrow u_2$ while the second factor is anti-symmetric.
Therefore only the following form, up to interchanges of roles of
$(d,d',d'')$,
 \bea
 &&(d\g_{m_1n_1p_1} d)(d'\g_{m_2n_2p_2} d')(d''\g_{m_3n_3p_3} d'')
 (d\g_{m_4n_4p_4} d')(d\g_{m_5n_5p_5} d'')(d'\g_{m_6n_6p_6}
 d'')\nn\\
 &&(\lb\g^{m_1n_1p_1}\,r)(\lb\g^{m_2n_2p_2}\,r)(\lb\g^{m_3n_3p_3}\,r)
 (\lb\g^{m_4n_4p_4}\,r)(\lb\g^{m_5n_5p_5}\,r)(\lb\g^{m_6n_6p_6}\,r)
 \eea
 is potentially non-vanishing. Taking the steps that are similar to
 those in section 2.2.1, one gets
 \bea
 &&(r\g^{rsq}\lb)(r\g^{r's'q'}\lb)(r\g^{r''s''q''}\lb)\nn\\
 &&(r\g^{u_4}\g^r \l)(\lb \g_{u_4}\g^{r'}\l)
  (r\g^{u_5}\g^s \l)(\lb \g_{u_5}\g^{r'}\l)
  (r\g^{u_4}\g^{s'} \l)(\lb \g_{u_4}\g^{s''}\l)\nn\\
 \sim &&(\lb\l)^3 (\lb\l)(r \g^{rsq}\lb) (r \g^{r's'q'}\lb)
              (r\g^{r''s''q''}\lb)\nn\\
    && (r\g^{r'}\g^r \l)(r\g^{r''}\g^s \l)(r\g^{s''}\g^{s'} \l)
 \eea
As in section 2.2.1, the factor, $(r\g^{r'}\g^r \l)$ can be replaced
by $\eta^{r'r}(\l r)$ and the expression vanishes.
The third term in \rf{3loop} is where the regulator ${\cal N}_2$
provides 33 $d$-zero modes. The six $b$-ghosts should provide 11
remaining $d$'s and the vertex operators 4. One gets
 \bea
 && \int [d\l][d\lb][dr] d^{16}\th \prod_{I=1}^3
[dw^I] [d\bar w^I] [ds^I] d^{16} d^I \nn\\
 &&e^{-\lb\l-r\th-\wb w+sd}
 \nn\\
&& \left[\fr{(\lb\g^{mnp}\,r)(d\g_{mnp}\,d)}
    {192(\lb\l)^2}\right]^{5}
    \left(\fr{\Pi^{m'} (\lb\g_{m'} d)}{2(\lb\l)}
       -\fr{(r\g_{m'n'p'}r)(\lb\g^{m'}d)N^{n'p'}}{16(\lb\l)^3}
            \right)      \nn\\
 &&[d_{\a_1}W^{\a_1}(z_1)][d_{\a_2}W^{\a_2}(z_2)]
 [d_{\a_3}W^{\a_3}(z_3)][ d_{\a_4}W^{\a_4}(z_4)]\nn\\
 \sim &&\int [d\l][d\lb][dr] d^{16}\th \prod_{I=1}^3
[dw^I] [d\bar w^I] [ds^I] d^{16} d^I \nn\\
 && e^{-\lb\l-r\th-\wb w } (sd)^{33}
 \nn\\
&& \left[\fr{(\lb\g^{mnp}\,r)(d\g_{mnp}\,d)}
    {192(\lb\l)^2}\right]^{5}(\lb\g^{m'}d)
    \left(\fr{\Pi^{m'}}{2(\lb\l)}
       -\fr{(r\g_{m'n'p'}r)N^{n'p'}}{16(\lb\l)^3}
            \right)      \nn\\
 &&[d_{\a_1}W^{\a_1}(z_1)][d_{\a_2}W^{\a_2}(z_2)]
 [d_{\a_3}W^{\a_3}(z_3)][ d_{\a_4}W^{\a_4}(z_4)]
 \eea
Performing $s$-integration, one gets
 \bea
 \sim && \int [d\l][d\lb][dr] d^{16}\th \prod_{I=1}^3
[dw^I][d\bar w^I]  d^{16} d^I \nn\\
  &&\left[\fr{}{}\right. (\l \g^r)_{\a_1}(\l \g^s)_{\a_2}(\l \g^q)_{\a_3}
   (\g_{rsq})_{\a_4\a_5}\;\e^{\a_1...\a_5\r_1...\r_{11}}
   d_{\r_1}...d_{\r_{11}} \nn\\
 &&\;\;(\l \g^{r'})_{\a_ 1'}(\l \g^{s'})_{\a_2'}(\l \g^{q'})_{\a_3'}
   (\g_{r's'q'})_{\a_4'\a_5'}\;\e^{\a_1'...\a_5'\r_1'...\r_{11}'}
   d'_{\r_1'}...d'_{\r_{11}'} \nn\\
 &&\;\;(\l \g^{r''})_{\a_ 1''}(\l \g^{s''})_{\a_2''}(\l \g^{q''})_{\a_3''}
   (\g_{r''s''q''})_{\a_4''\a_5''}\;\e^{\a_1''...\a_5''\r_1''...\r_{11}''}
   d''_{\r_1''}...d''_{\r_{11}''}\left.\fr{}{}\right] \nn\\
 && \left[\fr{(\lb\g^{mnp}\,r)(d\g_{mnp}\,d)}
    {192(\lb\l)^2}\right]^{5}(\lb\g^{m'}d)
    \left(\fr{\Pi^{m'}}{2(\lb\l)}
       -\fr{(r\g_{m'n'p'}r)N^{n'p'}}{16(\lb\l)^3}
            \right)      \nn\\
 &&[d_{\b_1}W^{\b_1}(z_1)][d_{\b_2}W^{\b_2}(z_2)]
 [d_{\b_3}W^{\b_3}(z_3)][ d_{\b_4}W^{\b_4}(z_4)]
 \label{3group}
 \eea
Let us focus on the following factors in \rf{3group},
 \bea
 &&\left[{(\lb\g^{mnp}\,r)(d\g_{mnp}\,d)}
    \right]^{5}(\lb\g^{m'}d)
 \label{dd5}
 \eea
With the experience gained in the analysis of $(d\g^\3 d)^6$ above,
it is not difficult to see that the only potentially non-vanishing
forms are
 \bea
  &&(d\g_{m_1n_1p_1}\,d)(d'\g_{m_2n_2p_2}\,d')
 (d\g_{m_3n_3p_3}\,d')
 (d\g_{m_4n_4p_4}\,d'')(d'\g_{m_5n_5p_5}\,d'')\nn\\
 &&(d\g_{m_1n_1p_1}\,d)(d'\g_{m_2n_2p_2}\,d')(d''\g_{m_3n_3p_3}\,d'')
 (d\g_{m_4n_4p_4}\,d') (d\g_{m_5n_5p_5}\,d'')
 \label{dd5nonzero}
 \eea
and the ones with $(d,d',d'')$ interchanged. As we illustrate below,
these terms vanish. Put (\ref{3group}) into three groups,
 \bea \mbox{I}:
&&\left[\fr{}{}\right. (\l \g^r)_{\a_1}(\l \g^s)_{\a_2}(\l
\g^q)_{\a_3}
   (\g_{rsq})_{\a_4\a_5}\;
   \nn\\
 &&\;\;(\l \g^{r'})_{\a_ 1'}(\l \g^{s'})_{\a_2'}(\l \g^{q'})_{\a_3'}
   (\g_{r's'q'})_{\a_4'\a_5'}\;
   \nn\\
 &&\;\;(\l \g^{r''})_{\a_ 1''}(\l \g^{s''})_{\a_2''}(\l \g^{q''})_{\a_3''}
   (\g_{r''s''q''})_{\a_4''\a_5''}\;
   \left.\fr{}{}\right], \nn\\
II: &&
(\g_{m_1n_1p_1})^{\k_1\k_2}(\g_{m_2n_2p_2})^{\k_3\k_4}(\g_{m_3n_3p_3})^{\k_5\k_6}
    (\g_{m_4n_4p_4})^{\k_7\k_8}(\g_{m_5n_5p_5})^{\k_9\k_{10}}\nn\\
     && (\lb_{\s_1}(\g^{m_1n_1p_1})^{\s_1\s_2}\,r_{\s_2})
     (\lb_{\s_3}(\g^{m_2n_2p_2})^{\s_3\s_4}\,r_{\s_4})
     (\lb_{\s_5}(\g^{m_3n_3p_3})^{\s_5\s_6}\,r_{\s_6})\nn\\
      && (\lb_{\s_7}(\g^{m_4n_4p_4})^{\s_7\s_8}\,r_{\s_8})
     (\lb_{\s_9}(\g^{m_5n_5p_5})^{\s_9\s_{10}}\,r_{\s_{10}}),
      \nn\\
 III: &&(\lb\g^{m'} )^\b[W^{\b_1}(z_1)][W^{\b_2}(z_2)]
 [W^{\b_3}(z_3)][W^{\b_4}(z_4)]
 \label{3groupdd5}
 \eea
 Let us take the first term of \rf{dd5nonzero},
 \bea
  &&(d\g_{m_1n_1p_1}\,d)(d'\g_{m_2n_2p_2}\,d')(d\g_{m_3n_3p_3}\,d')
 (d\g_{m_4n_4p_4}\,d'')(d'\g_{m_5n_5p_5}\,d'')
 \eea
The steps are similar to the case of the $(d\g^\3 d)^6$. After the
$[ds]$-, $[dd]$- integrations and Fierzing, one gets an expression
that contains
 \bea
  &&(r\g^{rsq}\lb)(r\g^{r's'q'}\lb)
 (r\g^{u_3}\g^r \l)(\lb \g_{u_3}\g^{r'}\l)\nn\\
 \sim && (\lb \l) (r\g^{rsq}\lb)(r\g^{r's'q'}\lb)
 (r\g^{r'}\g^r \l)
 \eea
The last factor yields $\eta^{r'r}$ rendering the expression to
vanish.

 \subsection{Aisaka-Berkovits regulator}

In the previous section, we have exhibited all the techniques that
we will use in this section. A proper regularization of the
$\fr{1}{\l\lb}$-singularities for three- and higher- loops was
outlined in \cite{Berkovits:2006vi}. The resulting formula is
complex. Setting apart the gauge invariance issue, Aisaka and
Berkovits have proposed a relatively simple regulator that is quoted
in \rf{thlr}. We take \rf{thlr} and work out a few contributions in
detail.

 First we survey the amount of terms that need to be
computed. One simplifying feature is that the $d$- and $[ds]$-
integrations should remove many of the terms, if not the majority.
Based on inspection of the $d$-integration, one concludes that only
the terms in the square bracket of \rf{3loop} are potentially
non-zero. In particular we narrow down to the last two
terms.\footnote{In addition, there are the extra contributions that
have been indicated in one of the footnotes in section 3.1.} The
zero mode integrations now become
 \bea
 \sim && \int [d\l][d\lb][dr] d^{16}\th \prod_{I=1}^3
[dw^I][d\bar w^I]  d^{16} d^I
 \exp[-\sum_I^g (\bar{w}_I^\a w_{\a,I})]
      \exp(f^\a w_\a+g^\a d_\a+\bar{f}_\a \bar{w}^\a)
\nn\\
  &&\left[\fr{}{}\right. (\l \g^r)_{\a_1}(\l \g^s)_{\a_2}(\l \g^q)_{\a_3}
   (\g_{rsq})_{\a_4\a_5}\;\e^{\a_1...\a_5\r_1...\r_{11}}
   (d_{\r_1}-\bar{g}_{\r_1})...(d_{\r_{11}}-\bar{g}_{\r_{11}}) \nn\\
 &&\;\;(\l \g^{r'})_{\a_ 1'}(\l \g^{s'})_{\a_2'}(\l \g^{q'})_{\a_3'}
   (\g_{r's'q'})_{\a_4'\a_5'}\;\e^{\a_1'...\a_5'\r_1'...\r_{11}'}
   (d'_{\r_1'}-\bar{g}_{\r'_1})...(d'_{\r'_{11}}-\bar{g}_{\r'_{11}}) \nn\\
 &&\;\;(\l \g^{r''})_{\a_ 1''}(\l \g^{s''})_{\a_2''}(\l \g^{q''})_{\a_3''}
   (\g_{r''s''q''})_{\a_4''\a_5''}\;\e^{\a_1''...\a_5''\r_1''...\r_{11}''}
   (d''_{\r_1''}-\bar{g}_{\r_1''})...(d''_{\r_{11}''}-\bar{g}_{\r''_{11}}) \left.\fr{}{}\right] \nn\\
 &&\left\{ \fr{}{}\right.
  \left[\fr{(\lb\g^{mnp}\,r)(d\g_{mnp}\,d)}
    {192(\lb\l)^2}\right]^{5}(\lb\g_{m'} d)
    \left(\fr{\Pi^{m'} }{2(\lb\l)}
       -\fr{(r\g_{m'n'p'}r)N^{n'p'}}{16(\lb\l)^3}
            \right)      \nn\\
 &&\quad\quad[d_{\a_1}W^{\a_1}(z_1)][d_{\a_2}W^{\a_2}(z_2)]
 [d_{\a_3}W^{\a_3}(z_3)][ d_{\a_4}W^{\a_4}(z_4)]\nn\\
 &&+\left[\fr{(\lb\g^{mnp}\,r)(d\g_{mnp}\,d)}
    {192(\lb\l)^2}\right]^{6}
     [\pa\th^{\a_1} A_{\a_1}+\Pi^{m_1} A_{m_1}
   +\fr12N^{m_1n_1}{\cal F}_{m_1n_1}] \nn\\
 &&\quad\quad [d_{\a_2}W^{\a_2}(z_2)]
 [d_{\a_3}W^{\a_3}(z_3)][ d_{\a_4}W^{\a_4}(z_4)]
 \left.\fr{}{}\right\}
 \label{ggbdgd6}
 \eea
Since $g^\a$ and $\bar{g}_\a$ are grassmanian, there can be maximum
eleven factors for each. Also since the measure contains
$d^{11}g,d^{11}\bar{g}$, there must be precisely eleven factors of
$g^\a, \bar{g}_\a$ respectively. The break-up of the contributions
goes as follows. The factor, $\exp(g^\a d_\a)$, contributes eleven
$d$-zero modes. The square bracket contributes twenty two $d$-zero
modes (and eleven $\bar{g}$'s). The last two factors then must
provide the remaining fifteen. The factor of $\exp(g^\a d_\a)$ and
the square bracket will produce terms of the form
 \bea
 && (\bar{g}_{\r_1}...\bar{g}_{\r_m}d_{\r_{m+1}}...d_{\r_{11}})
  ({g}^{\s_{1}}...{g}^{\s_{M}}{d}_{\s_{1}}...{d}_{\s_{M}})\nn\\
  && (\bar{g}_{\r'_1}...\bar{g}_{\r'_n}d'_{\r'_{n+1}}...d'_{\r'_{11}})
  ({g}^{\s'_{1}}...{g}^{\s'_{N}}d'_{\s'_{1}}...d'_{\s'_{N}})\nn\\
   && (\bar{g}_{\r''_1}...\bar{g}_{\r''_l}d''_{\r''_{l+1}}...d''_{\r''_{11}})
  ({g}^{\s''_{1}}...{g}^{\s''_{L}}d''_{\s''_{1}}...d''_{\s''_{L}})
 \label{ggb}
  \eea
with $m+n+l=11$ and $M+N+L=11$ as can be ssen by considering the
number of $\bar{g}$'s. Let us follow the following line of logic.
Suppose that \rf{ggb} contains less than eleven factors for some of
$(d,d',d'')$. The deficit must be compensated by the terms in the
square bracket in \rf{ggbdgd6}.
 The integrations over $dg$ and $d\bar{g}$ are
carried out with the following Lorentz invariant measures,
 \[
  \left[ d{g} \right] = \fr{1}{(\l\lb)^3}(\l\g^m)_{\a_1}(\l\g^n)_{\a_2}(\l\g^p)_{\a_3}
   (\g_{mnp})_{\a_4\a_5}\e^{\a_1...\a_{5}\d_1...\d_{11}}
   \pa^g_{\d_1}...\pa^g_{\d_{11}}
 \]
 \[
 \left[ d\bar{g} \right](\l\g^m)_{\a_1}(\l\g^n)_{\a_2}(\l\g^p)_{\a_3}
   (\g_{mnp})_{\a_4\a_5} = (\l\lb)^3\e_{\a_1...\a_{5}\d_1...\d_{11}}
   \pa_{\bar{g}}^{\d_1}...\pa_{\bar{g}}^{\d_{11}}
 \]
After carrying out the $g$-integration, one gets, among other
factors,
 \bea
 &&\e^{\a_1...\a_5\r_1...\r_{11}}
  \;\e^{\a_1'...\a_5'\r_1'...\r_{11}'}
 \; \e^{\a_1''...\a_5''\r_1''...\r_{11}''}
 \;\;\;\d_{\s_1...\s_M\s_1'...\s_N'\s_1''...\s_L''}
      ^{\r_1...\r_m\r_1'...\r_n'\r_1''...\r_l''}\nn\\
 &&
 (d_{\r_{m+1}}...d_{\r_{11}}d_{\s_1}...d_{\s_M})
  (d'_{\r_{n+1}'}...d'_{\r_{11}'}d'_{\s_1'}...d'_{\s_N'})
 (d''_{\r_{l+1}''}...d''_{\r_{11}''}d''_{\s_1''}...d''_{\s_L''})
 \label{delta-epsilon}
 \eea
 where the $\e$'s in the first line come from the $[ds]$-measure.
Because of the grassmanian nature of the $(d,d',d'')$ which are
dummy variables, the permutation in
$\d_{\s_1...\s_M\s_1'...\s_N'\s_1''...\s_L''}
      ^{\r_1...\r_m\r_1'...\r_n'\r_1''...\r_l''}$ can be entirely
 dropped. In all subsequent calculations we only consider
 \be
 \d_{\s_1}^{\r_1}\d_{\s_2}^{\r_2}\cdots \d_{\s_L''}^{\r_l''},
 \label{deltaorder}
 \ee
i.e., the term that preserves the exact order of the upper and
 the lower indices as indicated
 in $\d_{\s_1...\s_M\s_1'...\s_N'\s_1''...\s_L''}
      ^{\r_1...\r_m\r_1'...\r_n'\r_1''...\r_l''}$.
  The integration measure provides 33 to the 48 $d$-zero
 modes needed. We take the first term in (\ref{ggbdgd6}), which
 contains $(d\g^\3 d)^5$, for
 detailed analysis.
 The $d_0^I$-deficit must be compensated for by the factors of
the same kind in
 \bea
 && \left[\fr{(\lb\g^{mnp}\,r)(d\g_{mnp}\,d)}
    {192(\lb\l)^2}\right]^{5}(\lb\g_{m'} d)
    \left(\fr{\Pi^{m'} }{2(\lb\l)}
       -\fr{(r\g_{m'n'p'}r)N^{n'p'}}{16(\lb\l)^3}
            \right)      \nn\\
 &&[d_{\a_1}W^{\a_1}(z_1)][d_{\a_2}W^{\a_2}(z_2)]
 [d_{\a_3}W^{\a_3}(z_3)][ d_{\a_4}W^{\a_4}(z_4)]
 \label{d15}
 \eea
   Depending on how the fifteen deficit
$d$'s are distributed, the terms that arrive by expanding $(d\g
d)^5$ can be classified as
 \bea
 (n_d,n_{d'},n_{d''})=&& (15,0,0),(14,1,0),(13,2,0),(13,1,1),
 (12,3,0),(12,2,1),(11,4,0),\nn\\
 && (11,3,1),(11,2,2),(10,5,0),
 (10,4,1),(10,3,2),(9,6,0),(9,5,1),\nn\\
 &&(9,4,2),(9,3,3),(8,7,0),(8,6,1),(8,5,2),(8,4,3),(7,7,1),\nn\\
 &&(7,6,2),(7,5,3),
 (7,4,4),(6,6,3),(6,5,4),(5,5,5)
 \eea
 where we have assumed, $n_d\geq n_{d'}\geq n_{d''} $,  without
  loss of generality. It is because the cases otherwise will only change the
  irrelevant overall factors.
The cases with $n_d=15,...,12$ trivially vanish for the following
reason. In such cases, the expressions contain
 \bea
 \e_{...\k_1\k_2\k_3\k_4...}(\g_{m_1n_1p_1})^{\k_1\k_2}(\lb \g^{m_1n_1p_1} r)
                            (\g_{m_2n_2p_2})^{\k_3\k_3}(\lb \g^{m_2n_2p_2} r)
 \eea
This equation identically vanishes as one can see by interchanging
$(\g_{m_1n_1p_1})^{\k_1\k_2}(\lb \g^{m_1n_1p_1} r)$ and
$(\g_{m_2n_2p_2})^{\k_3\k_3}(\lb \g^{m_2n_2p_2} r)$. Also the term,
$(n_d,n_{d'},n_{d''})=(5,5,5)$, becomes essentially the same as the
case considered in the previous section. Therefore the terms that
require more careful inspection are
 \bea
 (n_d,n_{d'},n_{d''})= &&(11,4,0),(11,3,1),(11,2,2),
   (10,5,0),(10,4,1),(10,3,2),(9,6,0),\nn\\
 && (9,5,1),(9,4,2),(9,3,3),(8,7,0),(8,6,1),(8,5,2),(8,4,3),\nn\\
 &&(7,7,1),(7,6,2),(7,5,3),
 (7,4,4),(6,6,3),(6,5,4) \label{dgd5tobechecked}
 \eea
Each of these terms has sub-cases depending on how $(d,d',d'')$ is
 distributed over the factors that come from the vertex operators,
 $(\lb \g^{m'} d)(dW)(dW)(dW)(dW)$.

We now select a few example and maximally simplify them. After the
$[ds]$- and $[dd]$- integrations, a given contribution becomes an
equation where the variables, ($\l$'s, $\lb$'s, $r$'s,
$\g^{m_1...m_j}$'s) get contracted among themselves and/or with the
SYM fields. It gets simplified by applying the Fierz identities and
the pure spinor constraints. Often in the middle of manipulations,
one encounters the following form of factors,
 \bea
 (...\g^{p}\g^mr)(...\g_p\, r)
 \eea
Applying a gamma matrix identity it becomes,
 \bea
 (\chi \g^{p}\g^mr)(\xi\g_p r)= 2(\chi \,r)(\xi\g^m\, r)
                       -(\chi'\g^p\,r)(\xi\g_p\, r)
                       \label{chhixi}
 \eea
 where $\chi'\equiv \chi\g^{m}$.
In the examples below, we focus on the contributions coming from the
first part. One of the reasons, which we elaborate on in the
Appendix C and a footnote below, is that the contributions
associated with the second term tend to vanish for one reason or
another. In the approach that we are taking to simplify, it has
something to with the fact that the expression,
 \bea
 (\chi' \g^{\m\n\r} \xi)  (r\g_{\m\n\r} r),
 \eea
which results from applying one of the Fierz identity, does not
admit the obvious inverse Fierzing because of the changed index
structure. One can force the inverse Fierzing but then procedure
introduces $\g^0$'s. As a matter of fact, what we amount to show in
the appendix is that all the structures that do not contain a single
factor of $(\chi'\g^p\,r)(\xi\g_p\, r)$ drop out due to various
constraints and identities.

One of the differences between the two-loop regulator, \rf{tlr}, and
the three-loop regulator, \rf{thlr}, is that the three-loop
regulator introduces more mixing between terms through \rf{ggb}.
Comparison of the cases of $(n_d,n_{d'},n_{d''})=(5,5,5)$ and
$(n_d,n_{d'},n_{d''})=(6,5,4)$ may illustrate the point. The case,
(5,5,5), is essentially the same as the case that has been analyzed
in the previous section. When the $\a$-indices in \rf{3groupdd5} get
contract with the $\k$-indices, the indices of different numbers of
primes are not allowed to contract. It is partially allowed in the
latter case because (\ref{deltaorder}) takes the form of
 \bea
 && \d_{\k_1........\k_{6}}^{\a_1...\a_5\r_m}
 \quad
 \d_{\k_1'...\k_5'\r_m}^{\a_1'...\a_5'\r_n'}
 \quad
 \d_{\k_1''...\k_4''\r_n'}^{\a_1''.....\;\a_5''}
 \eea
The cases listed in (\ref{dgd5tobechecked}) allow various degrees of
mixing. Once there is a sufficient amount of mixing, the
contribution generates the same final result although the terms with
different level of mixing produce different looking expressions in
the intermediate steps. For the $(d\g^\3 d)^5$ case
 \bea
 && \left[{(\lb\g^{mnp}\,r)(d\g_{mnp}\,d)}
   \right]^{5}  \nn\\
 &&[d_{\a_1}W^{\a_1}(z_1)][d_{\a_2}W^{\a_2}(z_2)]
 [d_{\a_3}W^{\a_3}(z_3)][ d_{\a_4}W^{\a_4}(z_4)]
 \label{exemplary}
 \eea
Eq.\rf{exemplary} comes multiplied with $(\lb\g^{m'} d)
    {(r\g_{m'n'p'}r)N^{n'p'}}$ which can rewritten as
 \bea
 -8 (r \g_{m'})(r\l)+4(r\g_{m'}\g^\m \l)(r\g_\m w)
 \label{tailfactor}
 \eea
 after use of one of the Fierz identity.
 Because the \rf{3loopresult} contains
  one factor of $(\l r)$, eq.\rf{tailfactor} reduces to
 \bea
 -4(r \g^\m\g_{m'} \l)(r\g_\m w)
 \eea
 We have worked out several exemplary and random cases\footnote{The analysis
 still requires a large amount of algebra because each case has many sub-cases.}
  and found that
 they all reduce the following form
 \bea
  (\l r)
 (\l \g^{t_1}\g^{t_2}r)[r W][\l \g^{t_3}W]
 [r \g^{t_1}\g^{m'} W][r \g^{t_2}\g^{t_3}W]
 \label{3loopresult}
  \eea
 where the factor $(\l\lb)^6$ has been omitted.
 The index, $m'$, appears in (\ref{ggbdgd6}). Although we did not
  run a complete analysis, the result has been obtained from
 what we believe a few generic forms that could cover many different
 contributions. At some point of algebra, those apparent differences
 do not matter too much thanks to the anti-symmetry,
 \rf{anti-sym}, and various Fierz identities.

One aspect of the computation deserves a comment. In Appendix C, it
is discussed for the two-loop case that the terms of the form
$(...\g^p\,r )(...\g_p\, r)$ tend to have pairwise cancelations. We
show here that for the three-loop case as well they have a stronger
tendency to vanish. A given expression gets simplified by repeatedly
applying various Fierz identities and the pure spinor constraints.
 At the final stage of the process, typically a term of the
 following form arrives
 \bea
 (\lb\l)^6 (\l r)(\l \g^{r'}\g^{s} r)
 (W \g^{m'}\g^{s}\g^{r}\g^{r'}\,r) (r \g^{r}\g^{q'} W)
 (\l \g^{q'}W)(r W) \label{illu}
 \eea
 When one moves $\g^s$
in $(W \g^{m'}\g^{s}\g^{r}\g^{r'}\,r)$ to the right, one generates a
few terms of the form given in (\ref{3loopresult}). Moving it to the
far right and focusing on the two middle factors, one gets
 \bea
 && (\l \g^{r'}\g^{s} r)(W \g^{m'}\g^{r}\g^{r'}\g^{s}\,r)\nn\\
 \sim && (\l \g^{r'}\g^{0} r)(W \g^{m'}\g^{r}\g^{r'}\g^{0}\,r)
 \eea
where in the second line we have used a Fierz identity that is
derived in Appendix C. Since there is a factor of $(\l r)$ in
\rf{illu}, one can freely shift the location of $\g^{r'}$ as
 \bea
 \sim && (\l \g^{0}\g^{r'} r)(W \g^{m'}\g^{r}\g^{0}\g^{r'}\,r)
 \eea
 Application of the Fierz identity one more time brings another
 factor of $(\l r)$, hence the expression vanishes. In Appendix C,
It is shown that each contribution in the three-loop can be put into
the form of $(...\g^p\, r )(...\g_p\, r)$.
 Together with the the two-loop case it
may be taken as another indication that, with \rf{thlr}, the overall
coefficient of (\ref{3loopresult}) might vanish. If it is indeed
zero, the role of a gauge invariant regulator will be simply to
replace with by a non-zero number, as mentioned previously.

We take (\ref{3loopresult}) as a three-loop analogue of
(\ref{tlfinal}). It is proposed based on the tendency that has been
observed in all of the cases analyzed so far: the amplitude
structures such as (\ref{tlfinal}) and (\ref{3loopresult}) reside in
the {\em individual} terms that are generated by the different
$d_0$-structures and the terms that come via an expansion of the
regulator. The powerful antisymmetry made it possible to run a
generic analysis because many apparently different-looking terms
will correspond to different contractions that are allowed by the
anti-symmetry. As for the case of $(d\g^\3 d)^6$, without any
operator product among the non-zero modes, all of our attempts to
get an analogue of (\ref{3loopresult}) have led to vanishing
results.

\section{Conclusion}

After reviewing the one-loop and directly computing the two-loop, we
have tackled the three-loop amplitude. For the three-loop, we have
used the regulator recently proposed by Aisaka and Berkovits. Based
on a few indications, we have discussed the possibility that the AB
regulator might lead to a vanishing result. Putting the issue aside,
we have argued that it is possible to acquire the three-loop
amplitude. The result has been presented in (\ref{3loopresult}).
Strictly speaking, the result is conjectural in the sense that not
all the contributions are explicitly evaluated. We still believe
that the result is likely to be confirmed by future analyses.

To fully verify (\ref{3loopresult}), there are several things that
need to be done. One obvious thing is to work out the extra
contribution that was mentioned in \rf{gomez}. It does not pose any
new difficulty or subtlety: it is just additional algebra of the
same kind. (Although not recorded, a few terms were analyzed and
shown to vanish. The analysis thereof is by no means extensive.) To
our view, the task of the highest priority is an improved
understanding of the three-loop regulator. As pointed out in the
section 2 (and also in later sections), there are no contributions
from any of the higher order $\fr{1}{\l\lb}$-terms in ${\cal
A}_{3-loop}$ that come from the six $b$-insertions and have made the
two-loop regulator invalid as a three-loop regulator. It would be
more natural if the ${\cal A}_{3-loop}$ received some contributions
from those higher order terms. Checking whether the AB regulator
yields a vanishing amplitude by direct computation will be an
extremely tedious task. It would be far better if one could rely on
some sort of sophisticated argument. Also it might be worth trying
to come up with a new regulator that has this feature. A partial
check can be provided by fully evaluating (\ref{3loopresult}) to get
a momentum space expression. One can see whether the resulting
momentum space expression has some expected symmetry such as
cyclicity. One may compare its field theory limit with the
corresponding SYM three-loop computation.

 Although the role of a new regulator (or a gauge invariant
regulator, in case a new regulator with a different gauge fixing is
not easy to come by) may be simply to change the overall numerical
coefficient of (\ref{3loopresult}), we cannot entirely exclude the
possibility that it might introduce a new term that is different
 from (\ref{3loopresult}). At most it will be a few more terms if any.
  (The contribution of \rf{gomez} should also be
  checked against this possibility.) Checking more examples will be
  useful to strengthen the result, (\ref{3loopresult}). Once a new regulator
   or a gauge invariant regulator is used, some of
the contributions could come from the terms in higher powers in
$\fr1{\lb \l}$ than those in (\ref{ggbdgd6}). A gauge invariant
regulator will include more diverse structures in the pure spinor
field, $\l$ and its conjugate. The computation using a new regulator
or a gauge invariant regulator -which is one future direction- is
expected to be more complex. We expect the techniques and procedures
presented in this work to be useful in future computations. Progress
on these issues will be reported.

\vspace{.5in}
\ni {\bf Acknowledgements}\\
I thank Y. Aisaka, N. Berkovits, C. Mafra and P. van Hove for their
communications throughout the project. I especially thank H. Kawai
for arranging my trips to Kyoto University, Okayama Institute of
Quantum Physics and RIKEN. I am also grateful for their hospitality
during my visits.  I acknowledge the international visitor's program
of KEK and thank Y. Kitazawa for his hospitality during my visit.
Part of the work was carried out during my stay at Asia Pacific
Center for Theoretical Physics and I thank Y. Kim for his
hospitality.

\newpage

\renewcommand{\theequation}{A.\arabic{equation}}
 \setcounter{equation}{0}
  \section*{Appendix A: Notations}

Our index conventions are
 \bea
 && \m,a,m\; \mbox{etc} : \mbox{10D vector indices} \nn\\
 &&\a,\b\; \mbox{etc}: \mbox{10D spinor indices}
 \eea
 As for the gamma matrices, we distinguish the 32 by 32 gamma
 matrices, $\G^m$, from the 16 by 16 ones, $\g^m$,
\[
\G^m=\left(
           \begin{array}{cc}
        0  &  (\g^m)_{\a\b}        \\
        (\g^m)^{\a\b}  &   0      \\
   \end{array}
          \right)\;\;,\;\;
\]
 For example, $\G^{mn}$ takes the form of
\[
\G^{mn}=\left(
           \begin{array}{cc}
        (\g^{mn})_{\a}{}^{\b}   &   0      \\
        0  &   (\g^{mn})^{\a}{}_{\b}       \\
   \end{array}
          \right)\;\;,\;\;
\]
 in this representation.
 When taking the transpose in the 32-component it is convenient to
 use an explicit representation of $\G^0$. Without loss of
 generality one can choose
\[
\G^0=\left(
           \begin{array}{cc}
        0  &  1        \\
        -1  &   0      \\
   \end{array}
          \right)\;\;,\;\;
\]
When applying the 32-component Fierzing, we embed the 16-component
spinors in 32-component ones
 \bea
 \l_u&= \left(
           \begin{array}{c}
        \l^\a        \\
         0      \\
   \end{array}
          \right)\quad,\quad
           \l_d &=\left(
           \begin{array}{c}
        0       \\
         \l^\a      \\
   \end{array}
          \right)\nn\\
          r_u&=\left(
           \begin{array}{c}
        r_\a        \\
         0      \\
   \end{array}
          \right)\quad,\quad
           r_d &=\left(
           \begin{array}{c}
        0       \\
         r_\a      \\
   \end{array}
          \right)\nn\\
    W_u&= \left(
           \begin{array}{c}
        W^\a        \\
         0      \\
   \end{array}
          \right)\quad,\quad
           W_d &=\left(
           \begin{array}{c}
        0       \\
         W^\a      \\
   \end{array}
          \right)
 \eea
and similarly for other fields.
The 16 by 16 gamma matrices satisfy
 \bea
 \eta_{mn}\g^m_{\a(\b} \g^n_{\g\d)}=0
 \eea
 We often write $d^I$ instead of $d_0^I$ for the simplicity of notation.
In the abuse of the notation, we often use
 \bea
 d^{I=1}&\equiv & d \nn\\
 d^{I=2}& \equiv & d'\nn\\
 d^{I=3}& \equiv & d''
 \eea
  A dotted
equality, ``$\doteq$", represents an equality, ``=",
 up to an overall numerical factor,
 \bea
  ``\doteq"\;\mbox{denotes}\;  ``="
 \mbox{up to an overall numerical factor}
 \eea
 In some equations, the symbol, ``$\sim$", is used instead.
The symbol, $\oplus$, indicates
 \bea
 \oplus:\; \mbox{relative numerical
coefficients are not being recorded precisely}
 \eea
For example, $A \oplus B$ means $A+B$ with the relative numerical
coefficients between $A$ and $B$ inaccurate.

\renewcommand{\theequation}{B.\arabic{equation}}
 \setcounter{equation}{0}
  \section*{Appendix B: identities
   }

For 32 by 32 gamma matrices, one can show that
 \bea
&& \G^a\G_a=10,\quad \G^a \G^\m \G_a=-8\G^\m,\quad \G^a \G^{\m\n}
\G_a=6\G^{\m\n}
  \nn\\
 &&\G^a \G^{\m\n\r} \G_a=-4\G^{\m\n\r},\quad
  \G^a\G^{\m\n\r\s} \G_a=2\G^{\m\n\r\s}, \quad
 \G^a \G^{\m\n\r\s\d} \G_a=0\nn\\
 &&\G^a \G^{\m\n\r\s\d\k} \G_a=-2\G^{\m\n\r\s\d\k},\quad
 \G^a \G^{\m\n\r\s\d\k\z} \G_a=4\G^{\m\n\r\s\d\k\z}\nn\\
 &&\G^{ab}\G_{ab}=-90,\quad
  \G^{ab} \G^\m \G_{ab}=-54\G^\m,\quad \G^{ab} \G
 ^{\m\n} \G_{ab}=-26\G^{\m\n}\nn\\
 && \G^{ab} \G^{\m\n\r} \G_{ab}=-6\G
 ^{\m\n\r},\quad\G^{ab}
 \G^{\m\n\r\s} \G_{ab}=6\G
 ^{\m\n\r\s},\quad \G^{ab} \G^{\m\n\r\s\d} \G
 _{ab}=10 \G^{\m\n\r\s\d},\quad \nn\\
 &&\G^{ab} \G^{\m\n\r\s\d\k} \G
 _{ab}=6\G^{\m\n\r\s\d\k}\nn\\
 &&  \G^{abc}\G_{abc}=-720,\quad
  \G^{abc} \G^\m \G_{abc}=288\G^\m,\quad \G^{abc} \G^{\m\n} \G
 _{abc}=-48\G^{\m\n}
 \nn\\
 && \G^{abc} \G^{\m\n\r} \G_{abc}=-48\G^{\m\n\r},\quad \G^{abc}
 \G^{\m\n\r\s} \G_{abc}=48\G^{\m\n\r\s},\quad \G^{abc} \G^{\m\n\r\s\d}
\G_{abc}=0
 \nn\\
 &&  \G^{abc} \G
 ^{\m\n\r\s\d\k} \G_{abc}=-48\G^{\m\n\r\s\d\k}
\nn\\
 &&\G^{abcd}\G_{abcd}=5040,\quad
   \G^{abcd} \G^\m \G
 _{abcd}=1008\G^\m,\quad \G^{abcd} \G^{\m\n} \G_{abcd}=-336\G^{\m\n}
 \nn\\
 &&\G^{abcd} \G^{\m\n\r} \G_{abcd}=-336\G^{\m\n\r},\quad \G^{abcd}
 \G^{\m\n\r\s} \G
 _{abcd}=48\G^{\m\n\r\s}\nn\\
 && \G^{abcd} \G^{\m\n\r\s\d} \G_{abcd}=240 \G
 ^{\m\n\r\s\d},\quad
   \G^{abcd} \G^{\m\n\r\s\d\k} \G_{abcd}=48\G^{\m\n\r\s\d\k}\nn\\
 &&
 \G^{abcde}\G_{abcde}=6\cdot 5040,\quad
   \G^{abcde}\G^\m\G_{abcde}=0,\quad \G^{abcde}\G^{\m\n}\G_{abcde}=-3360\G^{\m\n},
 \quad  \nn\\
 &&\G^{abcde}\G^{\m\n\r}\G_{abcde}=0,\quad
   \G^{abcde}\G^{\m\n\r\s}\G_{abcde}=1440\G^{\m\n\r\s}\nn\\
   &&
 \G^{abcde}\G^{\m\n\r\s\d}\G_{abcde}=0
 \eea
They also satisfy\footnote{ Many of them can be found in
\cite{Candelas:1984yd} where a few of the coefficients are
erroneous. }
 \bea
 &&[\G _{m},\G^{r}]=2\G_{m}{}^{r}\quad,\quad
    \{\G _{m},\G^{r}\}=2\d_{m}{}^{r} \nn\\
 && \{\G _{mn},\G^{r}\}=2\G_{mn}{}^{r} \quad,\quad
    [\G _{mn},\G
    ^{r}]=-4\d_{[m}^{r}\G_{n]} \nn\\
 && [\G _{mnp},\G^{r}]=2\G_{mnp}{}^{r}\quad,\quad
    \{\G
     _{mnp},\G^{r}\}=6\,\d_{[m}^{r}\G_{np]}\nn\\
 && \{\G
  _{mnpq},\G^{r}\}=2\G_{mnpq}{}^{r} \quad,\quad
     [\G _{mnpq},\G^{r}]=-8\d_{[m}^{r}\G
     _{npq]}\nn\\
 && [\G _{mnpqk},\G^{r}]=2\G_{mnpqk}{}^{r} \quad,\quad
 \{\G _{mnpqk},\G^{r}\}=10\,\d_{[m}^{r}\G_{npqk]} \nn\\
 && \{\G
  _{mn},\G^{rs}\}=2\G_{mn}{}^{rs}-4\d_{[mn]}^{rs} \quad,\quad
   [\G _{mn},\G^{rs}]=-8\d_{[m}^{[r} \G_{n]}{}^{s]} \nn\\
 &&\{\G
 _{mnp}\,,\G^{rs}\}=2\G_{mnp}{}^{rs}-12\d_{[mn}^{rs}\G_{p]}\quad,\quad
   [\G _{mnp}\,,\G^{rs}]=12\d_{[m}^{[r}\G_{np]}{}^{s]}
 \nn\\
 && \{\G_{mnpq}\,,\G
 ^{rs}\}=2\G_{mnpq}{}^{rs}-24\d_{[mn}^{rs}\G_{pq]}
 \quad,\quad  [\G _{mnpq}\,,\G^{rs}]=-16\d_{[m}^{[r}\G
 _{npq]}{}^{s]}
 \nn\\
 &&\{\G_{mnpqk}\,,\G^{rs}\}=2\G_{mnpqk}{}^{rs}-40\d_{[mn}^{rs}\G_{pqk]}
 \quad,\quad  [\G
  _{mnpqk}\,,\G^{rs}]=20\d_{[m}^{[r}\G_{npqk]}{}^{s]}
 \nn\\
 &&[\G _{mnp},\G
 ^{rst}]=2\G_{mnp}{}^{rst}-36\;\d_{[mn}^{[rs}\G_{p]}{}^{t]}\,\quad
  \{\G _{mnp},\G^{rst}\}=18\d_{[m}^{[r}\G_{np]}{}^{rs]}-12\d_{[mnp]}^{rst}
  \nn\\
 &&\{\G
 _{mnpq}\,,\G^{rst}\}=2\G_{mnpq}{}^{rst}-72\d_{[mn}^{rs}\G_{pq]}{}^{t]}
 \quad\quad\nn\\
 &&[\G_{mnpq},\G^{rst}]=-24\d_{[m}^{[r}\G_{npq]}{}^{st]}+48\;\d_{[mnp}^{rst}\G_{q]}
   \nn\\
 &&\{\G
 _{mnpq}\,,\G^{rstu}\}=2\G_{mnpq}{}^{rstu}-144\d_{[mn}^{rs}\G_{pq]}{}^{tu]}
   +48\d_{[mnpq]}^{rstu}\quad,\quad\nn\\
  && [\G_{mnpq},\G^{rstu}]=-32\d_{[m}^{[r}\G_{npq}{}^{stu]}
                             -64\d_{[mnp}^{[rst}\G_{q}{}^{u]}
   \nn\\
 && \{\G_{mnpqk},\G
 ^{rst}\}=30\d_{[m}^{[r}\G_{npqk}{}^{st]}-120\;\d_{[mnp}^{[rst}\G_{qk]}
   \nn\\
 &&[\G_{mnpqk},\G^{rst}]=2\G
 _{mnpqk}{}^{rst}-120\;\d_{[mn}^{[rs}\G_{pqk]}{}^{t]}
   \nn\\
 && \{\G_{mnpqk},\G^{rstu}\}=2\G
 _{mnpqk}{}^{rstu}-240\;\d_{[mn}^{[rs}\G_{pqk}{}^{tu]}
   +240\;\d_{[mnpq}^{[rstu}\G_{k]}\nn\\
  && [\G_{mnpqk},\G^{rstu}]=40\;\d_{[m}^{[r}\G_{npqk}{}^{stu]}
   -480\;\d_{[mnp}^{[rst}\G
   _{qk]}{}^{u]}\nn\\
  && \{\G_{mnpqk},\G^{rstuw}\}=50\;\d_{[m}^{[r}\G_{npqk}{}^{stuw]}
  -1200\;\d_{[mnp}^{[rst}\G_{qk}{}^{uw]}
   +240\;\d_{mnpqk}^{rstuw}\nn\\
 && [\G
 _{mnpqk},\G^{rstuw}]=2\G_{mnpqk}{}^{rstuw}-400\;\d_{[mn}^{[rs}\G_{pqk}{}^{tuw]}
   +1200\;\d_{[mnpq}^{[rstu}\G_{k]}{}^{w]}\nn\\
  && \{\G_{mnpqkl},\G^{r}\}=2\G
  _{mnpqkl}{}^{r},\quad
  [\G_{mnpqkl},\G^{r}]=-12\;\d_{[m}^{r}\G_{npqkl]}\nn\\
 &&\{\G_{mnpqkl},\G^{rs}\}=2\G_{mnpqkl}{}^{rs}-60\;\d_{[mn}^{[rs}\G
 _{pqkl]},\quad
   [\G_{mnpqkl},\G^{rs}]=-24\;\d_{[m}^{[r}\G_{npqkl]}{}^{s]}
    \nn\\
 &&\{\G_{mnpqkl},\G^{rst}\}=2\G_{mnpqkl}{}^{rst}-180\;\d_{[mn}^{[rs}\G_{pqkl]}{}^{t]}
   \nn\\
 && [\G
 _{mnpqkl},\G^{rst}]=-36\d_{[m}^{[r}\G_{npqkl}{}^{st]}
          +240\;\d_{[mnp}^{rst}\G_{qkl]}
   \nn\\
 && \{\G_{mnpqkl},\G^{rstu}\}=2\G_{mnpqkl}{}^{rstu}-360\;\d_{[mn}^{[rs}\G
 _{pqkl}{}^{tu]}
   +720\;\d_{[mnpq}^{rstu}\G_{kl]}\nn\\
&& [\G_{mnpqkl},\G^{rstu}]=-48\;\d_{[m}^{[r}\G
 _{npqkl}{}^{stu]}
   +960\;\d_{[mnp}^{[rst}\G_{qkl]}^{u]}\nn\\
 && \{\G_{mnpqkl},\G^{rstuw}\}=-600\;\d_{[mn}^{[rs}\G_{pqkl}{}^{tuw]}
   +3600\;\d_{[mnpq}^{[rstu}\G_{kl]}{}^{w]}\nn\\
  && [\G_{mnpqkl},\G^{rstuw}]=-60\;\d_{[m}^{[r}\G_{npqkl}{}^{stuw]}
  +2400\;\d_{[mnp}^{[rst}\G_{qkl}{}^{uw]}
   -1440\;\d_{[mnpqk}^{rstuw}\G_{l]}\nn\
 \eea
In some manipulations, we used
 \bea
 \G^{\m_1...\m_4}&=&\G^{\m_4}...\G^{\m_1}-\eta^{\m_1\m_2}\G^{\m_4}\G
 ^{\m_3}
 -2\G^{\m_4}\eta^{\m_3[\m_2}\G^{\m_1]}+3\eta^{\m_4[\m_1}\G
 ^{\m_2\m_3]}\nn\\
                &=&\G^{\m_1}...\G^{\m_4}-\eta^{\m_1\m_2}\G
                ^{\m_3}\G^{\m_4}
 +2\eta^{\m_3[\m_1}\G^{\m_2]}\G^{\m_4}-3\eta^{\m_4[\m_1}\G
 ^{\m_2\m_3]}\nn\\
 \G^{\m_1...\m_3}&=&-\G^{\m_3}\G^{\m_2}\G^{\m_1}
       + \eta^{\m_1\m_2}\G^{\m_3}-2\eta^{\m_3[\m_1}\G^{\m_2]}\nn\\
       &=&\G^{\m_1}\G^{\m_2}\G^{\m_3}
       - \eta^{\m_1\m_2}\G^{\m_3}+2\eta^{\m_3[\m_1}\G^{\m_2]}\nn\\
 \G^{\m_1...\m_5}&=&\G^{\m_5}...\G^{\m_1}-\eta^{\m_1\m_2}\G^{\m_5}\G^{\m_4}\G^{\m_3}
 -2\G^{\m_5}\G^{\m_4}\eta^{\m_3[\m_2}\G^{\m_1]}\nn\\
 &&+3\G^{\m_5}\eta^{\m_4[\m_1}\G^{\m_2\m_3]}
 -4\eta^{\m_5[\m_1}\G^{\m_2\m_3\m_4]}
 \eea
 which are straightforward to verify.
  For commuting spinors, the Fierz rearrangement is
 \bea
 (\chi_1^\dagger M \chi_2)(\chi_3^\dagger N \chi_4)
 =\sum_I (\chi_1^\dagger O^I \chi_4)(\chi_3^\dagger NO^IM \chi_2)
 \fr{1}{\tr(O_I^2)}
 \eea
For anti-commuting spinors, there will be an overall minus sign. By
choosing $M,N$ and $\chi$'s appropriately one can derive the
16component Fierz relations. For example, with $M=\G^m,
N=\G^m,\chi_1=(1_\a ,0),\chi_2=(0,1_\b ),\chi_3=(1_\s
,0),\chi_4=(0,1_\d)$, one gets
 \bea
 \g^m_{\a\b}\g^m_{\d\s}=-\fr12 \g^m_{\a\d}\g^m_{\b\s}
   +\fr1{24}\g^{mnp}_{\a\d}\g^{mnp}_{\s\b} \label{gamma3}
 \eea
 Taking $M=\G^{mnpq}, N=\G^{mnpq}$, one
gets
 \bea
 (\g_{mnpq})_\a{}^{\b}(\g_{mnpq})_\s{}^{\d}=315\d_{\a}^\d\d_{\s}^\b
 +\fr{21}{2}(\g^{mn})_{\a}{}^\d (\g^{mn})_{\s}{}^\b
               +\fr18 (\g^{mnpq})_{\a}{}^\d (\g^{mnpq})_{\s}{}^\b
 \eea
 Taking $M=\G^{mnp}, N=\G^{mnp}$ with the same choice of
 $\chi$'s gives,\footnote{Compared with (62) of \cite{Alexandrov:2007pd}, there is
 overall minus sign
difference.}
 \bea
 \g_{\a\b}^{abc}\g_{\d\s}^{abc}=18\g_{\a\d}^m\g_{\s\b}^m
               +\fr12 \g_{\a\d}^{mnp}\g_{\s\b}^{mnp}
 \eea
 Taking either of the equation above and
anti-symmetrizing $(\b\s\d)$, one can easily show
 \bea
 \g^m_{\a(\b}\g^m_{\d\s)}=0
 \eea
Another useful identity given in \cite{Berkovits:2006vi} is
 \bea
 (\g^{\m\n})_\a{}^\d (\g^{\m\n})_\b{}^\s
 =-8 \d_\a^\s \d_\b^\d+4(\g^{\m})_{\a\b} (\g^{\m})^{\d\s}
  -2\d_\a^\d \d_\b^\s
 \eea
For $\tr(O_I^2)$, note that one can make the following replacements,
 \bea
 && \tr(1)\ra 32,\quad \tr((\G^{\2})^2)\ra -64,\quad
\tr((\G^{\4})^2)\ra
 768,\quad \tr((\G^{\6})^2)\ra -6!\cdot 32 \nn\\
 && \tr((\G^{\8})^2)\ra 8!\cdot 32,\quad \tr((\G^{(10)})^2)\ra
 -10!\cdot 32 \nn\\
 && \tr(\G^{\1})\ra 32,\quad \tr((\G^{\3})^2)\ra -3!\cdot 32,\quad
\tr((\G^{\5})^2)\ra
 5!\cdot 32,\quad \tr((\G^{\7})^2)\ra -7!\cdot 32 \nn\\
 && \tr((\G^{(9)})^2)\ra 9!\cdot 32
 \eea
 The following relations were used when a transpose of a 32 by 32 gamma
 matrix was taken,
 \bea
(\G^\m)^T&=& \G^0\G^\m \G^0\nn\\
 (\G^{\m\n})^T&=&\G^0 \G^{\m\n} \G^0 \nn\\
 (\G^{\m\n\r})^T&=&-\G^0 \G^{\m\n\r} \G
 ^0\nn\\
 (\G^{\m\n\r\s})^T&=&-\G^0 \G^{\m\n\r\s} \G^0\nn\\
 (\G^{\m_1...\m_5})^T&=&\G^0 \G^{\m_1...\m_5} \G^0
 \eea
The fact that $\l_u \G^{\m\n\r}\l_d=0$ can be seen, e.g., by using
the transpose relations given above.

\renewcommand{\theequation}{C.\arabic{equation}}
 \setcounter{equation}{0}
  \section*{Appendix C: Aisaka-Berkovits regulator, cont'd}

After the $[ds]$- and $[dd]$- integrations, a given contribution
becomes an equation where the variables, ($\l$'s, $\lb$'s, $r$'s,
$\g^{m_1...m_j}$'s) get contracted among themselves and/or with the
SYM fields. It gets simplified by applying the Fierz identities and
the pure spinor constraints. While doing so, one encounters the
following form of factors,
 \bea
 (...\g^{p}\g^mr)(...\g_p r)
 \eea
We write this as $(\chi \g^{p}\g^mr)(\xi\g_p r)$ which in turn can
be rewritten,
 \bea
 (\chi \g^{p}\g^mr)(\xi\g_p r)= 2(\chi \,r)(\xi\g^m\, r)
                       -(\chi'\g^p\,r)(\xi\g_p\, r)
                       \label{chhixiq}
 \eea
 where $\chi'\equiv \chi\g^{m}$. Let us focus on the second term.
The Fierz identity, $r_\a r_\b\sim (\g^{\m\n\r})_{\a\b}
(r\g_{\m\n\r}r)$, yields
 \bea
 (\chi\g^p\,r)(\xi\g_p\, r)\sim (\chi \g^{\m\n\r} \xi)  (r\g_{\m\n\r} r),
 \eea
The prime on $\chi'$ has been dropped. It does not admit the inverse
Fierzing right away because of the changed index structure. One can
``force" the inverse Fierzing by temporarily going to the
32-component notation with the first factor,
 \bea
 \sim &&(\chi_d \G^{\m\n\r} \xi_u)  (r\g_{\m\n\r} r)\nn\\
 \sim &&(  \xi_u (\G^{\m\n\r})^T \chi_d)  (r\g_{\m\n\r} r)\nn\\
 \sim && (  \xi_u \G^0\G^{\m\n\r}\G^0 \chi_d)  (r\g_{\m\n\r} r)
 \eea
 where in the third line one of the gamma matrix identity has been
 used. Moving back to the 16-component notation, one gets
 \bea
 \sim && (  \xi \g^0\g^{\m\n\r}\g^0 \chi)  (r\g_{\m\n\r} r)
 \eea
 The matrix, $\g^0$, is $\d^{\a\b}$ up to a sign. Its role is to
 bring back the index structure so now the inverse
 Fierzing can be carried out,
 \bea
  (\chi\g^p\,r)(\xi\g_p\, r)\sim  ( \chi \g^0 r)(  \xi \g^0 r)
  \label{if}
 \eea
Note that by forcing the inverse Fierzing, extra $\g^0$'s have been
introduced. As a matter of fact, what we amount to show below is
that all the structures that do not contain a single factor of
$(\chi'\g^p\,r)(\xi\g_p\, r)$ drop out due to various constraints
and identities. Eq.\rf{if} has an interesting implication. We
illustrate it with a specific example. In the two-loop analysis of
the $(d\g^\3 d)^5$-term, one typically encounters a term such as
 \bea
 (r\l)(\l \g^r W_\1)(\l \g^{s'} W_\3)(r \g^{s'}\g^{q'} W_\4)(r\g^{q'} \g^r W_\2)
 \eea
at some stage of the manipulation. Writing
 \bea
(r \g^{s'}\g^{q'} W_\4)=-(r \g^{q'} \g^{s'} W_\4)+2 (rW_\4)
 \eea
 we focus on the first term since the second term yields the desired
 two-loop expression, \rf{tlfinal}. With the other factors,
 the first term gives
 \bea
 (r\l)(\l \g^r W_\1)(\l \g^{s'} W_\3)(r \g^{q'}\g^{s'} W_\4)(r\g^{q'} \g^r W_\2)
 \eea
Substituting \rf{if}, one gets
 \bea
 (r\l)(\l \g^r W_\1)(\l \g^{s'} W_\3)(r \g^{0}\g^{s'} W_\4)(r\g^{0} \g^r W_\2)
 \eea
which looks more ``uniform" compared with \rf{tlfinal}. This term
gets canceled by a term that is present due to the permutation of
the external states,
 \bea
 (r\l)(\l \g^r W_\3)(\l \g^{s'} W_\1)(r \g^{0}\g^{s'} W_\2)(r\g^{0} \g^r W_\4)
 \eea
as one can easily check by shifting around $W$'s and relabeling the
indices, $(r,s')$.

We come to the task of showing that the entire three-loop
contributions with the regulator \rf{thlr} can be put into the forms
that contain at least one factor of $(...\g^p \,r)(...\g^p\, r)$.
Mostly the analysis of $(d\g^\3 d)^5$-case is presented. The
analysis of $(d\g^\3 d)^6$-case is simpler. We have conducted
complete analysis examining all the sub-cases
 for the following cases,
 \bea
 (n_d,n_{d'},n_{d''})= &&(11,4,0),(11,3,1),(11,2,2),
   (10,5,0),\nn\\
   && (10,4,1),(10,3,2),(7,5,3),(6,5,4)
   \label{dgd5_complete}
 \eea
 and run partial checks on,
 \bea
 (n_d,n_{d'},n_{d''})= &&(8,4,3),(7,5,3)
 \label{dgd5_partial}
 \eea
 We illustrate with $(n_d,n_{d'},n_{d''})= (11,2,2)$. In appropriate places
 below, the word ``vanish" means ``vanish up to the terms that contain
 $(...\g^p\, r)(...\g_p\, r)$",
 \be
 {``vanish"} = {``vanish\;\; up\;\; to\;\; the\;\; terms\;\; that\;\; contain}\;
 (...\g^p\, r)(...\g_p\, r)"
 \ee
 Also a ``vanishing expression" means it is the form of $(...\g^p \,r)(...\g_p
 \,r)$. Frequent uses will be made of the identity, (\ref{fueproof2}).
 Because the analyses of the two terms in (\ref{fueproof2}) go similarly we present
 the contributions from the second term,
 \bea
  (\lb_d\G^{mnp}\,r_u)(\l_u \G^r \G_{mnp} \G^s\l_d)
 \Rightarrow (\lb\l)(\l_u \G^r \G^s r_u) \label{fueproof2-1}
  \eea
After the $g$- and $d$- integrations of the first term in
(\ref{ggbdgd6}), one gets
 \bea
 && \d_{\s_1...\s_{m-6}\;\;\s_1'.......\s_6'\;\s_7'...\s_{n+3}'\;\;\s_1''\;
           \;\;\s_2''\;\;\s_3''...\s_4''.....\s_{l+3}''}
           ^{\r_1...\r_{m-6}\r_{m-5}...\r_m\;\r_1'...\r_{n-3}'\r_{n-2}'\r_{n-1}'\r_n'\;
               \;\r_1''......\;\r_l''}\nn\\
 &&(\e_{\r_{m+1}...{\r_{11}}{\s_1}...{\s_{m-6}\k_1...\k_{11}}})
  (\e_{\r_{n+1}'...{\r_{11}'}{\s_1'}...{\s_{n+3}'}\k_1'\k_2'})
 (\e_{\r_{l+1}''...{\r_{11}''}{\s_1''}...{\s_{l+3}''}{\k_1''}\k_2''})
 \eea
Using the Kronecker-delta in front, one gets
 \bea
 (\e_{\r_{m+1}...{\r_{11}}{\r_1}...{\r_{m-6}\k_1...\k_{11}}})
  (\e_{\r_{n+1}'...{\r_{11}'}{\r_{m-5}}...\r_m\r_1'...{\r_{n-3}'}\k_1'\k_2'})
 (\e_{\r_{l+1}''...{\r_{11}''}{\r_{n-2}'}...\r_n'\r_1''...{\r_{l}''}{\k_1''}\k_2''})
 \eea
It can be contracted with other $\e$'s that come from the $[ds]$-
integration measure to yield
 \bea
 && \d_{\k_1.\;.\;.\;.\;.\;.\;.\;.\;.\;.\k_{11}}^{\a_1...\a_5\r_{m-5}...\r_m}
 \quad
 \d_{\k_1'\k_2'\r_{m-5}...\r_m}^{\a_1'...\a_5'\r_{n-2}'...\r_n'}
 \quad
 \d_{\k_1''\k_2''\r_{n-2}'...\r_n'}^{\a_1''.\;.\;.\;.\;.\;
        .\;.\;\a_5''}
        \label{11_2_2_threedelta}
 \eea
Fop this case not to vanish trivially, five $d_0$'s out of eleven
must be assigned to
 \bea
(\lb\g_m d)[dW(z_1)][dW(z_2)]
 [dW(z_3)][ dW(z_4)]
 \eea
 Otherwise two factors of a diagonal term, $(d^I\g^\3 d^I)(d^I\g^\3
 d^I)$, (no sum)
 will be present and the expression vanishes.
The factor, $(d \g_{mnp} d)^5$, in \rf{d15} takes the form of
 \bea
 ( \g_{m_1n_1p_1} )^{\k_1\k_2}( \g_{m_2n_2p_2} )^{\k_3\k_1'}
 ( \g_{m_3n_3p_3} )^{\k_4\k_2'}( \g_{m_4n_4p_4} )^{\k_5\k_1''}
 ( \g_{m_5n_5p_5} )^{\k_6\k_2''}
 \eea
We execute the third bundle of Kronecker-deltas in
\rf{11_2_2_threedelta} with the following choice,
 \bea
 \a_1''=\r_{n-2}',\quad \a_2''=\r_{n-1}',\quad\a_3''=\r_{n}',\quad
 \a_4''=\k_1'',\quad \a_5''=\k_2''
 \eea
They lead to
 \bea
 &&( \g_{m_1n_1p_1} )^{\k_1\k_2}( \g_{m_2n_2p_2} )^{\k_3\k_1'}
 ( \g_{m_3n_3p_3} )^{\k_4\k_2'}
 \left( \g_{m_4n_4p_4} \g^{r''s''q''}\g_{m_5n_5p_5} \right)^{\k_5\k_6}
 \nn\\
 &&\d_{\k_1.\;.\;.\;.\;.\;.\;.\;.\;.\;.\k_{11}}^{\a_1...\a_5\r_{m-5}...\r_m}
 \quad
 \d_{\k_1'\k_2'\r_{m-5}...\r_m}^{\a_1'...\a_5'\a_1''...\a_3''}
 \eea
The contractions dictated by
$\d_{\k_1.\;.\;.\;.\;.\;.\;.\;.\;.\;.\k_{11}}^{\a_1...\a_5\r_{m-5}...\r_m}$
can be classified into the following cases.\\
(i) When $(\a_1,...,\a_5)\in (\k_1,...,\k_6) $, one gets, e.g., $(\l
\g^r \g^{m_2n_2p_2}\g^{\Box} \l)\Tr\left(
\g^{rsq}\g^{m_1n_1p_1}\right)$ where $\Box$ denotes one of the
indices of $(r',s',q',r'',s'',q'')$. With other factors, it leads to
a
vanishing expression. \\
(ii) When four of $(\a_1,...,\a_5)$ contracts among
$(\k_1,...,\k_6)$, one can choose the contractions to generate a
factor, either of the previous type or of the form
 \bea
 (\l \g^r \g^{m_1n_1p_1}\g^{s} \l)\Tr\left(
\g^{rsq}\g^{m_4n_4p_4}\g^{r''s''q''}\g^{m_5n_5p_5}\right)
 \eea
which vanishes. \\
(iii)When $(\a_1,...,\a_5)$ contracts with three out of
$(\k_1,...,\k_6)$ and two $(\k_7,...,\k_{11})$, if the three $\k$'s
are $(\k_1,\k_2,\k_3)$, one gets $(\l \g^r \g^{m_2n_2p_2}\g^{\Box}
\l)\Tr\left( \g^{rsq}\g^{m_1n_1p_1}\right)$. If they are
$(\k_1,\k_2)$ and $\k_5$ ($\k_6$-case is similar), one gets
 \bea
 &&(\l \g^r \g^{m_1n_1p_1}\g^{s} \l)
 \left(\g^{rsq}\g^{m_4n_4p_4}\g^{r''s''q''}
    \g^{m_5n_5p_5}\right)^{\a_5\k_6}\nn\\
 &&(\g^{m_2n_2p_2})^{\k_3\k_1'}(\g^{m_3n_3p_3})^{\k_4\k_2'}\;\;
 \d_{\k_1'\k_2'\k_3\k_4\k_6\k_9...\k_{11}}
  ^{\a_1'..........\;\a_5'\a_1''...\a_3''}
 \eea
If there is any contraction between $(\a_1'',\a_1'',\a_3'')$ and
$(\k_1',\k_2',\k_3,\k_4)$, it is easy to see that the resulting
expression vanishes. The remaining case is
$(\a_1'',\a_1'',\a_3'')\in (\k_6\k_9...\k_{11})$. It leaves
 \bea
 (\g^{m_2n_2p_2})^{\k_3\k_1'}(\g^{m_3n_3p_3})^{\k_4\k_2'}\;\;
 \d_{\k_1'\k_2'\k_3\k_4\k}
  ^{\a_1'..........\;\a_5'}
 \eea
where $\k$ is one of the indices in $(\k_6,\k_9,...,\k_{11})$. The
contractions can be chosen to yield $(\l \g^{r'}
\g^{m_3n_3p_3}\g^{s'} \l)\Tr\left(\g^{r's'q'}\g^{m_2n_2p_2}\right)$
which vanishes. The other cases of three-$\k$ combinations can
similarly be
 shown to
vanish.\\
(iv)When the indices, $(\a_1,...,\a_5)$, contract with two out of
$(\k_1,...,\k_6)$ and three out of $(\k_7,...,\k_{11})$, one can
narrow down to the case where $(\a_1'',\a_2'',\a_3'')\in
(\k_5,\k_6,\k_{10},\k_{11})$ if the two $\k$'s are $(\k_1,\k_2)$.
Since that will leave $\d_{\k_1'\k_2'\k_3\k_4\;\k}
  ^{\a_1'..........\;\a_5'}$, this case vanishes too. The case where
 the two $\k$'s are $\k_1 (\mbox{or}\; \k_2)$ and
 $\k_3 (\mbox{or}\; \k_4)$ yields the following (or something similar)
 \bea
 &&(\l \g^r\g^{m_1n_1p_1})^{\k_2}(\l\g^s\g^{m_2n_2p_2})^{\k_1'}
 (\g^{m_3n_3p_3})^{\k_4\k_2'} \nn\\
 &&\left(\g^{m_4n_4p_4}\g^{r''s''q''}
    \g^{m_5n_5p_5}\right)^{\k_5\k_6}
    \d_{\k_1'\k_2'\k_2\k_4\k_5\k_6\k_{10}\k_{11}}
  ^{\a_1'..........\;\a_5'\a_1''...\a_3''}
 \eea
Any contraction between indices $(\k_1', \k_2',\k_1,\k_2)$ and
$(\a_1'', \a_2'',\a_3'')$ leads to zero: one is left with
 \bea
 (\l \g^r\g^{m_1n_1p_1})^{\k_2}(\l\g^s\g^{m_2n_2p_2})^{\k_1'}
 (\g^{m_3n_3p_3})^{\k_4\k_2'}  \d_{\k_1'\k_2'\k_2\k_4\k}
  ^{\a_1'..........\;\a_5'}
 \eea
which can be contracted to a vanishing expression. The
remaining cases can similarly be shown to vanish.\\
(v) When the indices, $(\a_1,...,\a_5)$ contract with one of
$(\k_1,...,\k_6)$ and four of $(\k_7,...,\k_{11})$, the case whose
vanishing is least trivial is when $(\a_1'',\a_2'',\a_3'')$ contract
with $\k_5,\k_6$ and one of $(\k_7,...,\k_{11})$, say, $\k_{11}$,
the case that results when $ (\k_7,...,\k_{10})\in (\a_1,...,\a_5)$.
Then the following factors are available
 \bea
 &&(\l \g^r\g^{m_1n_1p_1})^{\k_2}\left(\g^{m_2n_2p_2}\g^{r''s''q''}
    \g^{m_3n_3p_3}\right)^{\k_1'\k_2'}
    \d_{\k_1'\k_2'\k_2...\k_4}
  ^{\a_1'..........\;\a_5'}
 \eea
which vanishes.\\
 (vi) The analysis of $(\a_1,...,\a_5)\in (\k_7,...,\k_{11})$
goes similarly to that of (v).\\

Let us examine the $(n_d,n_{d'},n_{d''})= (6,5,4)$. In this case,
eq.\rf{delta-epsilon} becomes
\bea
 &&\e^{\a_1...\a_5\r_1...\r_{11}}
  \;\e^{\a_1'...\a_5'\r_1'...\r_{11}'}
 \; \e^{\a_1''...\a_5''\r_1''...\r_{11}''}
 \;\;\;\d_{\s_1...\s_{m-1}\s_1'\s_2'...\;\;\s_n'\;\;\s_1''\s_2''...\s_{l+1}''}
      ^{\r_1...\r_{m-1}\r_m\r_1'...\r_{n-1}'\r_{n}'\r_1''...\;\r_l''}\nn\\
 &&
 (d_{\r_{m+1}}...d_{\r_{11}}d_{\s_1}...d_{\s_{m-1}})
  (d'_{\r_{n+1}'}...d'_{\r_{11}'}d'_{\s_1'}...d'_{\s_n'})
 (d''_{\r_{l+1}''}...d''_{\r_{11}''}d''_{\s_1''}...d''_{\s_{l+1}''})
 \eea
 Carrying out the $d$-integrations, (\ref{delta-epsilon}) becomes
 \bea
 &&\e^{\a_1...\a_5\r_1...\r_{11}}
  \;\e^{\a_1'...\a_5'\r_1'...\r_{11}'}
 \; \e^{\a_1''...\a_5''\r_1''...\r_{11}''}
 \;\;\;\d_{\s_1...\s_{m-1}\s_1'\s_2'...\;\;\s_n'\;\;\s_1''\s_2''...\s_{l+1}''}
      ^{\r_1...\r_{m-1}\r_m\r_1'...\r_{n-1}'\r_{n}'\r_1''...\;\r_l''}\nn\\
 &&
 (\e_{\r_{m+1}...{\r_{11}}{\s_1}...{\s_{m-1}}\k_1...\k_6})
  (\e_{\r_{n+1}'...{\r_{11}'}\r_m{\s_1'}...{\s_n'}\k_1'...\k_5'})
 (\e_{\r_{l+1}''...{\r_{11}''}\r_{n'}{\s_1''}...{\s_{l+1}''}\k_1''...\k_4''})
 \label{ee}
 \eea
Using the Kronecker-delta in the manner indicated in
\rf{deltaorder}, one can choose the $\e$-contractions so as to
produce
 \bea
 && \d_{\k_1........\k_{6}}^{\a_1...\a_5\r_m}
 \quad
 \d_{\k_1'...\k_5'\r_m}^{\a_1'...\a_5'\r_n'}
 \quad
 \d_{\k_1''...\k_4''\r_n'}^{\a_1''.....\;\a_5''}
        \label{6_5_4_threedelta}
 \eea
We classify the cases based on the distribution of $d$'s over $(\lb
\g^m r)(dW)(dW)(dW)(dW)$:
 $\bullet\; 5d \in (\lb \g^m r)(dW)^4$
 \bea
 &&(d'\g_{m_1n_1p_1}\,d')(d'\g_{m_2n_2p_2}\,d)(d'\g_{m_3n_3p_3}\,d'')
 (d'\g_{m_4n_4p_4}\,d'') (d''\g_{m_5n_5p_5}\,d'')\nn\\
 &&(d'\g_{m_1n_1p_1}\,d')(d\g_{m_2n_2p_2}\,d'')(d'\g_{m_3n_3p_3}\,d'')
 (d'\g_{m_4n_4p_4}\,d'') (d'\g_{m_5n_5p_5}\,d'')\nn\\
 &&(d\g_{m_1n_1p_1}\,d')(d'\g_{m_2n_2p_2}\,d'')(d'\g_{m_3n_3p_3}\,d'')
 (d'\g_{m_4n_4p_4}\,d'') (d'\g_{m_5n_5p_5}\,d'')
 \label{1stexample}
 \eea
 $\bullet\; 4d,d' \in (\lb \g^m r)(dW)^4$
 \bea
 &&(d\g_{m_1n_1p_1}\,d)(d'\g_{m_2n_2p_2}\,d')(d'\g_{m_3n_3p_3}\,d'')
 (d'\g_{m_4n_4p_4}\,d'') (d''\g_{m_5n_5p_5}\,d'')\nn\\
 &&(d\g_{m_1n_1p_1}\,d)(d'\g_{m_2n_2p_2}\,d'')(d'\g_{m_3n_3p_3}\,d'')
 (d'\g_{m_4n_4p_4}\,d'') (d'\g_{m_5n_5p_5}\,d'')\nn\\
 &&(d\g_{m_1n_1p_1}\,d')(d\g_{m_2n_2p_2}\,d'')(d'\g_{m_3n_3p_3}\,d')
 (d'\g_{m_4n_4p_4}\,d'') (d''\g_{m_5n_5p_5}\,d'')\nn\\
 &&(d\g_{m_1n_1p_1}\,d')(d\g_{m_2n_2p_2}\,d')(d'\g_{m_3n_3p_3}\,d'')
 (d'\g_{m_4n_4p_4}\,d'') (d''\g_{m_5n_5p_5}\,d'')\nn\\
 &&(d\g_{m_1n_1p_1}\,d')(d\g_{m_2n_2p_2}\,d'')(d'\g_{m_3n_3p_3}\,d'')
 (d'\g_{m_4n_4p_4}\,d'') (d'\g_{m_5n_5p_5}\,d'')
 \eea
 $\bullet\; 4d,d'' \in (\lb \g^m r)(dW)^4$
 \bea
 &&(d\g_{m_1n_1p_1}\,d)(d'\g_{m_2n_2p_2}\,d')(d'\g_{m_3n_3p_3}\,d'')
 (d'\g_{m_4n_4p_4}\,d'') (d'\g_{m_5n_5p_5}\,d'')\nn\\
 &&(d\g_{m_1n_1p_1}\,d')(d\g_{m_2n_2p_2}\,d')(d'\g_{m_3n_3p_3}\,d')
 (d'\g_{m_4n_4p_4}\,d'') (d''\g_{m_5n_5p_5}\,d'')\nn\\
 &&(d\g_{m_1n_1p_1}\,d')(d\g_{m_2n_2p_2}\,d')(d'\g_{m_3n_3p_3}\,d'')
 (d'\g_{m_4n_4p_4}\,d'') (d'\g_{m_5n_5p_5}\,d'')
 \eea
 $\bullet\; 3d,2d' \in (\lb \g^m r)(dW)^4$
 \bea
 &&(d\g_{m_1n_1p_1}\,d)(d\g_{m_2n_2p_2}\,d'')(d'\g_{m_3n_3p_3}\,d')
 (d'\g_{m_4n_4p_4}\,d'') (d''\g_{m_5n_5p_5}\,d'')\nn\\
 &&(d\g_{m_1n_1p_1}\,d)(d\g_{m_2n_2p_2}\,d')(d'\g_{m_3n_3p_3}\,d'')
 (d'\g_{m_4n_4p_4}\,d'') (d''\g_{m_5n_5p_5}\,d'')\nn\\
 &&(d\g_{m_1n_1p_1}\,d)(d\g_{m_2n_2p_2}\,d'')(d'\g_{m_3n_3p_3}\,d'')
 (d'\g_{m_4n_4p_4}\,d'') (d'\g_{m_5n_5p_5}\,d'')\nn\\
 &&(d\g_{m_1n_1p_1}\,d')(d\g_{m_2n_2p_2}\,d'')(d\g_{m_3n_3p_3}\,d'')
 (d'\g_{m_4n_4p_4}\,d') (d''\g_{m_5n_5p_5}\,d'')\nn\\
 &&(d\g_{m_1n_1p_1}\,d')(d\g_{m_2n_2p_2}\,d')(d\g_{m_3n_3p_3}\,d'')
 (d'\g_{m_4n_4p_4}\,d'') (d''\g_{m_5n_5p_5}\,d'')\nn\\
 &&(d\g_{m_1n_1p_1}\,d')(d\g_{m_2n_2p_2}\,d'')(d\g_{m_3n_3p_3}\,d'')
 (d'\g_{m_4n_4p_4}\,d'') (d'\g_{m_5n_5p_5}\,d'')
 \label{2ndexample}
 \eea
 $\bullet\; 3d,d',d'' \in (\lb \g^m r)(dW)^4$
 \bea
 &&(d\g_{m_1n_1p_1}\,d)(d'\g_{m_2n_2p_2}\,d')(d\g_{m_3n_3p_3}\,d')
 (d'\g_{m_4n_4p_4}\,d'') (d''\g_{m_5n_5p_5}\,d'')\nn\\
 &&(d\g_{m_1n_1p_1}\,d)(d'\g_{m_2n_2p_2}\,d')(d'\g_{m_3n_3p_3}\,d'')
 (d'\g_{m_4n_4p_4}\,d'') (d\g_{m_5n_5p_5}\,d'')\nn\\
 &&(d\g_{m_1n_1p_1}\,d)(d\g_{m_2n_2p_2}\,d')(d'\g_{m_3n_3p_3}\,d'')
 (d'\g_{m_4n_4p_4}\,d'') (d'\g_{m_5n_5p_5}\,d'')\nn\\
 &&(d\g_{m_1n_1p_1}\,d')(d\g_{m_2n_2p_2}\,d')(d\g_{m_3n_3p_3}\,d'')
 (d'\g_{m_4n_4p_4}\,d') (d''\g_{m_5n_5p_5}\,d'')\nn\\
 &&(d\g_{m_1n_1p_1}\,d')(d\g_{m_2n_2p_2}\,d')(d\g_{m_3n_3p_3}\,d')
 (d'\g_{m_4n_4p_4}\,d'') (d''\g_{m_5n_5p_5}\,d'')\nn\\
 &&(d\g_{m_1n_1p_1}\,d')(d\g_{m_2n_2p_2}\,d')(d\g_{m_3n_3p_3}\,d'')
 (d'\g_{m_4n_4p_4}\,d'') (d'\g_{m_5n_5p_5}\,d'')
 \eea
  $\bullet\; 3d,2d'' \in (\lb \g^m r)(dW)^4$
 \bea
 &&(d\g_{m_1n_1p_1}\,d)(d\g_{m_2n_2p_2}\,d')(d'\g_{m_3n_3p_3}\,d')
 (d'\g_{m_4n_4p_4}\,d'') (d'\g_{m_5n_5p_5}\,d'')\nn\\
 &&(d\g_{m_1n_1p_1}\,d')(d\g_{m_2n_2p_2}\,d')(d\g_{m_3n_3p_3}\,d')
 (d'\g_{m_4n_4p_4}\,d') (d''\g_{m_5n_5p_5}\,d'')\nn\\
 &&(d\g_{m_1n_1p_1}\,d')(d\g_{m_2n_2p_2}\,d')(d\g_{m_3n_3p_3}\,d')
 (d'\g_{m_4n_4p_4}\,d'') (d'\g_{m_5n_5p_5}\,d'')
 \eea
 $\bullet\; 2d,3d' \in (\lb \g^m r)(dW)^4$
 \bea
 &&(d\g_{m_1n_1p_1}\,d)(d\g_{m_2n_2p_2}\,d'')(d\g_{m_3n_3p_3}\,d'')
 (d'\g_{m_4n_4p_4}\,d') (d''\g_{m_5n_5p_5}\,d'')\nn\\
 &&(d\g_{m_1n_1p_1}\,d)(d\g_{m_2n_2p_2}\,d')(d\g_{m_3n_3p_3}\,d'')
 (d'\g_{m_4n_4p_4}\,d'') (d''\g_{m_5n_5p_5}\,d'')\nn\\
 &&(d\g_{m_1n_1p_1}\,d)(d\g_{m_2n_2p_2}\,d'')(d\g_{m_3n_3p_3}\,d'')
 (d'\g_{m_4n_4p_4}\,d'') (d'\g_{m_5n_5p_5}\,d'')\nn\\
 &&(d\g_{m_1n_1p_1}\,d'')(d\g_{m_2n_2p_2}\,d'')(d\g_{m_3n_3p_3}\,d'')
 (d\g_{m_4n_4p_4}\,d'') (d'\g_{m_5n_5p_5}\,d')\nn\\
 &&(d\g_{m_1n_1p_1}\,d')(d\g_{m_2n_2p_2}\,d')(d\g_{m_3n_3p_3}\,d'')
 (d\g_{m_4n_4p_4}\,d'') (d''\g_{m_5n_5p_5}\,d'')\nn\\
 &&(d\g_{m_1n_1p_1}\,d')(d\g_{m_2n_2p_2}\,d'')(d\g_{m_3n_3p_3}\,d'')
 (d\g_{m_4n_4p_4}\,d'') (d'\g_{m_5n_5p_5}\,d'')
 \eea
 $\bullet\; 2d,2d',d'' \in (\lb \g^m r)(dW)^4$
 \bea
 &&(d\g_{m_1n_1p_1}\,d)(d\g_{m_2n_2p_2}\,d')(d\g_{m_3n_3p_3}\,d'')
 (d'\g_{m_4n_4p_4}\,d') (d''\g_{m_5n_5p_5}\,d'')\nn\\
 &&(d\g_{m_1n_1p_1}\,d)(d\g_{m_2n_2p_2}\,d')(d\g_{m_3n_3p_3}\,d')
 (d'\g_{m_4n_4p_4}\,d'') (d''\g_{m_5n_5p_5}\,d'')\nn\\
 &&(d\g_{m_1n_1p_1}\,d)(d\g_{m_2n_2p_2}\,d')(d\g_{m_3n_3p_3}\,d'')
 (d'\g_{m_4n_4p_4}\,d'') (d'\g_{m_5n_5p_5}\,d'')\nn\\
 &&(d\g_{m_1n_1p_1}\,d')(d\g_{m_2n_2p_2}\,d'')(d\g_{m_3n_3p_3}\,d'')
 (d\g_{m_4n_4p_4}\,d'') (d'\g_{m_5n_5p_5}\,d')\nn\\
 &&(d\g_{m_1n_1p_1}\,d')(d\g_{m_2n_2p_2}\,d')(d\g_{m_3n_3p_3}\,d')
 (d\g_{m_4n_4p_4}\,d'') (d''\g_{m_5n_5p_5}\,d'')\nn\\
 &&(d\g_{m_1n_1p_1}\,d')(d\g_{m_2n_2p_2}\,d')(d\g_{m_3n_3p_3}\,d'')
 (d\g_{m_4n_4p_4}\,d'') (d'\g_{m_5n_5p_5}\,d'')
 \label{3rdexample}
 \eea
  $\bullet\; 2d,d',2d'' \in (\lb \g^m r)(dW)^4$
 \bea
 &&(d\g_{m_1n_1p_1}\,d)(d\g_{m_2n_2p_2}\,d')(d\g_{m_3n_3p_3}\,d')
 (d'\g_{m_4n_4p_4}\,d') (d''\g_{m_5n_5p_5}\,d'')\nn\\
&&(d\g_{m_1n_1p_1}\,d)(d\g_{m_2n_2p_2}\,d')(d\g_{m_3n_3p_3}\,d'')
 (d'\g_{m_4n_4p_4}\,d') (d'\g_{m_5n_5p_5}\,d'')\nn\\
 &&(d\g_{m_1n_1p_1}\,d)(d\g_{m_2n_2p_2}\,d')(d\g_{m_3n_3p_3}\,d')
 (d'\g_{m_4n_4p_4}\,d'') (d'\g_{m_5n_5p_5}\,d'')\nn\\
&&(d'\g_{m_1n_1p_1}\,d')(d\g_{m_2n_2p_2}\,d')(d\g_{m_3n_3p_3}\,d')
 (d\g_{m_4n_4p_4}\,d'') (d\g_{m_5n_5p_5}\,d'')\nn\\
 &&(d\g_{m_1n_1p_1}\,d')(d\g_{m_2n_2p_2}\,d')(d\g_{m_3n_3p_3}\,d')
 (d\g_{m_4n_4p_4}\,d') (d''\g_{m_5n_5p_5}\,d'')\nn\\
 &&(d\g_{m_1n_1p_1}\,d')(d\g_{m_2n_2p_2}\,d')(d\g_{m_3n_3p_3}\,d')
 (d\g_{m_4n_4p_4}\,d'') (d'\g_{m_5n_5p_5}\,d'')
 \eea
  $\bullet\; 2d,3d'' \in (\lb \g^m r)(dW)^4$
 \bea
 &&(d\g_{m_1n_1p_1}\,d)(d'\g_{m_2n_2p_2}\,d')(d\g_{m_3n_3p_3}\,d')
 (d\g_{m_4n_4p_4}\,d') (d'\g_{m_5n_5p_5}\,d'')\nn\\
 &&(d\g_{m_1n_1p_1}\,d')(d\g_{m_2n_2p_2}\,d')(d\g_{m_3n_3p_3}\,d')
 (d\g_{m_4n_4p_4}\,d'') (d'\g_{m_5n_5p_5}\,d')\nn\\
 &&(d\g_{m_1n_1p_1}\,d')(d\g_{m_2n_2p_2}\,d')(d\g_{m_3n_3p_3}\,d')
 (d\g_{m_4n_4p_4}\,d') (d'\g_{m_5n_5p_5}\,d'')
 \eea
 $\bullet\; d,4d' \in (\lb \g^m r)(dW)^4$
 \bea
 &&(d\g_{m_1n_1p_1}\,d)(d\g_{m_2n_2p_2}\,d')(d\g_{m_3n_3p_3}\,d'')
 (d\g_{m_4n_4p_4}\,d'') (d''\g_{m_5n_5p_5}\,d'')\nn\\
 &&(d\g_{m_1n_1p_1}\,d)(d\g_{m_2n_2p_2}\,d'')(d\g_{m_3n_3p_3}\,d'')
 (d\g_{m_4n_4p_4}\,d'') (d'\g_{m_5n_5p_5}\,d'')\nn\\
 &&(d\g_{m_1n_1p_1}\,d')(d\g_{m_2n_2p_2}\,d'')(d\g_{m_3n_3p_3}\,d'')
 (d\g_{m_4n_4p_4}\,d'') (d\g_{m_5n_5p_5}\,d'')
 \eea
 $\bullet\; d,3d',d'' \in (\lb \g^m r)(dW)^4$
 \bea
 &&(d\g_{m_1n_1p_1}\,d)(d\g_{m_2n_2p_2}\,d'')(d\g_{m_3n_3p_3}\,d'')
 (d\g_{m_4n_4p_4}\,d'') (d'\g_{m_5n_5p_5}\,d')\nn\\
 &&(d\g_{m_1n_1p_1}\,d)(d\g_{m_2n_2p_2}\,d')(d\g_{m_3n_3p_3}\,d')
 (d\g_{m_4n_4p_4}\,d'') (d''\g_{m_5n_5p_5}\,d'')\nn\\
 &&(d\g_{m_1n_1p_1}\,d)(d\g_{m_2n_2p_2}\,d')(d\g_{m_3n_3p_3}\,d'')
 (d\g_{m_4n_4p_4}\,d'') (d'\g_{m_5n_5p_5}\,d'')\nn\\
 &&(d\g_{m_1n_1p_1}\,d')(d\g_{m_2n_2p_2}\,d')(d\g_{m_3n_3p_3}\,d'')
 (d\g_{m_4n_4p_4}\,d'') (d\g_{m_5n_5p_5}\,d'')
 \eea
 $\bullet\; d,2d',2d'' \in (\lb \g^m r)(dW)^4$
 \bea
 &&(d\g_{m_1n_1p_1}\,d)(d\g_{m_2n_2p_2}\,d')(d\g_{m_3n_3p_3}\,d'')
 (d\g_{m_4n_4p_4}\,d'') (d'\g_{m_5n_5p_5}\,d')\nn\\
 &&(d\g_{m_1n_1p_1}\,d)(d\g_{m_2n_2p_2}\,d')(d\g_{m_3n_3p_3}\,d')
 (d\g_{m_4n_4p_4}\,d') (d''\g_{m_5n_5p_5}\,d'')\nn\\
 &&(d\g_{m_1n_1p_1}\,d')(d\g_{m_2n_2p_2}\,d')(d\g_{m_3n_3p_3}\,d')
 (d\g_{m_4n_4p_4}\,d'') (d\g_{m_5n_5p_5}\,d'')
 \eea
 $\bullet\; d,d',3d'' \in (\lb \g^m r)(dW)^4$
 \bea
 &&(d\g_{m_1n_1p_1}\,d)(d\g_{m_2n_2p_2}\,d')(d\g_{m_3n_3p_3}\,d')
 (d\g_{m_4n_4p_4}\,d'') (d'\g_{m_5n_5p_5}\,d')\nn\\
 &&(d\g_{m_1n_1p_1}\,d)(d\g_{m_2n_2p_2}\,d')(d\g_{m_3n_3p_3}\,d')
 (d\g_{m_4n_4p_4}\,d') (d'\g_{m_5n_5p_5}\,d'')\nn\\
 &&(d\g_{m_1n_1p_1}\,d')(d\g_{m_2n_2p_2}\,d')(d\g_{m_3n_3p_3}\,d')
 (d\g_{m_4n_4p_4}\,d') (d\g_{m_5n_5p_5}\,d'')
 \eea
 $\bullet\; d,4d'' \in (\lb \g^m r)(dW)^4$
 \bea
 &&(d\g_{m_1n_1p_1}\,d)(d\g_{m_2n_2p_2}\,d')(d\g_{m_3n_3p_3}\,d')
 (d\g_{m_4n_4p_4}\,d') (d'\g_{m_5n_5p_5}\,d')\nn\\
 &&(d\g_{m_1n_1p_1}\,d')(d\g_{m_2n_2p_2}\,d')(d\g_{m_3n_3p_3}\,d')
 (d\g_{m_4n_4p_4}\,d') (d\g_{m_5n_5p_5}\,d')
 \eea
 $\bullet\; 5d' \in (\lb \g^m r)(dW)^4$
 \bea
 &&(d\g_{m_1n_1p_1}\,d)(d\g_{m_2n_2p_2}\,d'')(d\g_{m_3n_3p_3}\,d'')
 (d\g_{m_4n_4p_4}\,d'') (d\g_{m_5n_5p_5}\,d'')
 \eea
 $\bullet\; 4d',d'' \in (\lb \g^m r)(dW)^4$
 \bea
 &&(d\g_{m_1n_1p_1}\,d)(d\g_{m_2n_2p_2}\,d')(d\g_{m_3n_3p_3}\,d'')
 (d\g_{m_4n_4p_4}\,d'') (d\g_{m_5n_5p_5}\,d'')
 \eea
  $\bullet\; 3d',2d'' \in (\lb \g^m r)(dW)^4$
 \bea
 &&(d\g_{m_1n_1p_1}\,d)(d\g_{m_2n_2p_2}\,d')(d\g_{m_3n_3p_3}\,d')
 (d\g_{m_4n_4p_4}\,d'') (d\g_{m_5n_5p_5}\,d'')
 \eea
  $\bullet\; 2d',3d'' \in (\lb \g^m r)(dW)^4$
 \bea
 &&(d\g_{m_1n_1p_1}\,d)(d\g_{m_2n_2p_2}\,d')(d\g_{m_3n_3p_3}\,d')
 (d\g_{m_4n_4p_4}\,d') (d\g_{m_5n_5p_5}\,d'')
 \eea
 $\bullet\; d',4d'' \in (\lb \g^m r)(dW)^4$
 \bea
 &&(d\g_{m_1n_1p_1}\,d)(d\g_{m_2n_2p_2}\,d')(d\g_{m_3n_3p_3}\,d')
 (d\g_{m_4n_4p_4}\,d') (d\g_{m_5n_5p_5}\,d')
 \eea
We have used short-hand notations. For example, the case where five
$d$'s of the six $d$'s (six since $n_d=6$) come from $(\lb \g^m
r)(dW)(dW)(dW)(dW)$ has been denoted by
 \bea
 5\,d's \in (\lb \g^m r)(dW)(dW)(dW)(dW)
 \eea
There are multiple ways to show that each of these terms vanishes.
We illustrate the computations with a few examples. Our first
example is the first term in (\ref{1stexample}),
 \bea
 (d'\g_{m_1n_1p_1}\,d')(d'\g_{m_2n_2p_2}\,d)(d'\g_{m_3n_3p_3}\,d'')
 (d'\g_{m_4n_4p_4}\,d'') (d''\g_{m_5n_5p_5}\,d'')
 \label{6_5_4_d_distribution}
 \eea
The $d$-integration yields
 \bea
 (\g_{m_1n_1p_1})^{\k_1'\k_2'}(\g_{m_2n_2p_2})^{\k_3'\k_1}
 (\g_{m_3n_3p_3})^{\k_4'\k_1''}
 (\g_{m_4n_4p_4})^{\k_5'\k_2''} (\g_{m_5n_5p_5})^{\k_3''\k_4''}
 \eea
The index contractions dictated by the third factor of
Kronecker-deltas in (\ref{6_5_4_threedelta}) can be chosen as
 \bea
 \a_1''=\k_1'',\quad \a_2''=\k_2'',\quad\a_3''=\r_{n}',\quad
 \a_4''=\k_3'',\quad \a_5''=\k_4'' \label{achoice}
 \eea
 It leads to
 \bea
 ( \g^{m_3n_3p_3}\g^{r''} \l)^{\k_4'}( \g^{m_4n_4p_4}\g^{s''} \l)^{\k_5'}
 \;\Tr\left(\g^{r''s''q''}\g^{m_5n_5p_5}\right)\;
 \d_{\k_1'...\k_5'\r_m}^{\a_1'...\a_5'\a_3''}
 \quad
 \eea
which always reduces to an vanishing expression. Our second example
is the sixth term in (\ref{2ndexample}):
 \bea
 &&(d\g_{m_1n_1p_1}\,d')(d\g_{m_2n_2p_2}\,d'')(d\g_{m_3n_3p_3}\,d'')
 (d'\g_{m_4n_4p_4}\,d'') (d'\g_{m_5n_5p_5}\,d'')
 \eea
 The $d$-integration followed by the same choice as (\ref{achoice})
 yields,
 \bea
 ( \g^{m_2n_2p_2}\g^{r''} \l)^{\k_2}( \g^{m_3n_3p_3}\g^{s''} \l)^{\k_3}
 \;\left(\g^{m_4n_4p_4}\g^{r''s''q''}\g^{m_5n_5p_5}\right)^{\k_2'\k_3'}\;
 \d_{\k_1...\k_5\k_6}^{\a_1...\a_5\r_m}
 \d_{\k_1'...\k_5'\r_m}^{\a_1'...\a_5'\a_3''}
 \quad
 \eea
 where only the relevant factors are recorded.
 As one can easily check, it always leads to a vanishing expression. Let us take the fourth term in
 (\ref{3rdexample}), the case where
 $2d,2d',d'' \in (\lb \g^m r)(dW)^4$, as the third example,
  \bea
 (d\g_{m_1n_1p_1}\,d')(d\g_{m_2n_2p_2}\,d'')(d\g_{m_3n_3p_3}\,d'')
 (d\g_{m_4n_4p_4}\,d'') (d'\g_{m_5n_5p_5}\,d')
  \eea
  The steps similar to those above yield,
 \bea
 ( \g^{m_2n_2p_2}\g^{r''} \l)^{\k_2}
 \;\left(\g^{m_3n_3p_3}\g^{r''s''q''}\g^{m_4n_4p_4}\right)^{\k_3\k_4}\;
 \d_{\k_1...\k_5\k_6}^{\a_1...\a_5\r_m}
 \d_{\k_1'...\k_5'\r_m}^{\a_1'...\a_5'\a_3''}
 \quad
 \eea
 It again leads to a vanishing expression.\\

\newpage


\begin{thebibliography}{20}




\bibitem{Polchinski:1995mt}
  J.~Polchinski,
  ``Dirichlet-Branes and Ramond-Ramond Charges,''
  Phys.\ Rev.\ Lett.\  {\bf 75}, 4724 (1995)
  [arXiv:hep-th/9510017];   ``Lectures on D-branes,''
  arXiv:hep-th/9611050


\bibitem{pol}
  J. Polchinski,
  String theory, vol 1,2, Cambridge


\bibitem{Johnson:2000ch}
  C.~V.~Johnson,
  ``D-brane primer,''
  arXiv:hep-th/0007170.






\bibitem{Banks:1996vh}
  T.~Banks, W.~Fischler, S.~H.~Shenker and L.~Susskind,
  ``M theory as a matrix model: A conjecture,''
  Phys.\ Rev.\  D {\bf 55} (1997) 5112
  [arXiv:hep-th/9610043].



\bibitem{Ishibashi:1996xs}
  N.~Ishibashi, H.~Kawai, Y.~Kitazawa and A.~Tsuchiya,
  ``A large-N reduced model as superstring,''
  Nucl.\ Phys.\  B {\bf 498}, 467 (1997)
  [arXiv:hep-th/9612115].







\bibitem{Maldacena:1997re}
  J.~M.~Maldacena,
  ``The large N limit of superconformal field theories and supergravity,''
  Adv.\ Theor.\ Math.\ Phys.\  {\bf 2}, 231 (1998)
  [Int.\ J.\ Theor.\ Phys.\  {\bf 38}, 1113 (1999)]
  [arXiv:hep-th/9711200].


\bibitem{Gubser:1998bc}
  S.~S.~Gubser, I.~R.~Klebanov and A.~M.~Polyakov,
  ``Gauge theory correlators from non-critical string theory,''
  Phys.\ Lett.\  B {\bf 428}, 105 (1998)
  [arXiv:hep-th/9802109].

\bibitem{Witten:1998qj}
  E.~Witten,
  ``Anti-de Sitter space and holography,''
  Adv.\ Theor.\ Math.\ Phys.\  {\bf 2}, 253 (1998)
  [arXiv:hep-th/9802150].


\bibitem{Aharony:1999ti}
  O.~Aharony, S.~S.~Gubser, J.~M.~Maldacena, H.~Ooguri and Y.~Oz,
  ``Large N field theories, string theory and gravity,''
  Phys.\ Rept.\  {\bf 323}, 183 (2000)
  [arXiv:hep-th/9905111].












%
\bibitem{Kawai:2007ek}
  H.~Kawai and T.~Suyama,
  ``AdS/CFT Correspondence as a Consequence of Scale Invariance,''
  arXiv:0706.1163 [hep-th],
;
  ``Some Implications of Perturbative Approach to AdS/CFT Correspondence,''
  Nucl.\ Phys.\  B {\bf 794}, 1 (2008)
  [arXiv:0708.2463 [hep-th]].







\bibitem{Park:2001bm}
  I.~Y.~Park,
  ``Strong coupling limit of open strings: Born-Infeld analysis,''
  Phys.\ Rev.\ D {\bf 64}, 081901 (2001)
  [arXiv:hep-th/0106078].




\bibitem{Nielsen:1973qs}
  H.~B.~Nielsen and P.~Olesen,
  ``Local Field Theory Of The Dual String,''
  Nucl.\ Phys.\  B {\bf 57} (1973) 367.

\bibitem{Gibbons:2000hf}
  G.~W.~Gibbons, K.~Hori and P.~Yi,
  ``String fluid from unstable D-branes,''
  Nucl.\ Phys.\  B {\bf 596}, 136 (2001)
  [arXiv:hep-th/0009061].

\bibitem{Sen:2000kd}
  A.~Sen,
  ``Fundamental strings in open string theory at the tachyonic vacuum,''
  J.\ Math.\ Phys.\  {\bf 42}, 2844 (2001)
  [arXiv:hep-th/0010240].












\bibitem{Park:2007mc}
  I.~Y.~Park,
  ``Scattering on D3-branes,''
  Phys.\ Lett.\  B {\bf 660}, 583 (2008)
  [arXiv:0708.3452 [hep-th]]
;
``One loop scattering on D-branes,'' Eur. Phys. J. C {\bf 62}: 783
 (2009), arXiv:0801.0218 [hep-th]



\bibitem{Park:2008fp}
  I.~Y.~Park,
  ``Open string engineering of D-brane geometry,''
  JHEP {\bf 0808}, 026 (2008)
  [arXiv:0806.3330 [hep-th]].




\bibitem{Park:2009ki}
  I.~Y.~Park,
  ``Geometric counter-vertex for open string scattering on D-branes,''
  arXiv:0902.1279 [hep-th], to appear in EPJC.





































\bibitem{Zheng:2002ji}
  Z.~J.~Zheng, J.~B.~Wu and C.~J.~Zhu,
  ``Two-loop superstrings in hyperelliptic language. III: The four-particle
  amplitude,''
  Nucl.\ Phys.\  B {\bf 663}, 95 (2003)
  [arXiv:hep-th/0212219].


\bibitem{D'Hoker:2005jc}
  E.~D'Hoker and D.~H.~Phong,
  ``Two-Loop Superstrings VI: Non-Renormalization Theorems and the 4-Point
  Function,''
  Nucl.\ Phys.\  B {\bf 715}, 3 (2005)
  [arXiv:hep-th/0501197].



















\bibitem{Berkovits:2000fe}
  N.~Berkovits,
  ``Super-Poincare covariant quantization of the superstring,''
  JHEP {\bf 0004}, 018 (2000)
  [arXiv:hep-th/0001035]
;
  ``ICTP lectures on covariant quantization of the superstring,''
  arXiv:hep-th/0209059
;
  ``Pure spinor formalism as an N = 2 topological string,''
  JHEP {\bf 0510}, 089 (2005)
  [arXiv:hep-th/0509120].


\bibitem{Siegel:1985xj}
  W.~Siegel,
  ``Classical Superstring Mechanics,''
  Nucl.\ Phys.\  B {\bf 263}, 93 (1986).


\bibitem{Berkovits:2002qx}
  N.~Berkovits and O.~Chandia,
  ``Massive superstring vertex operator in D = 10 superspace,''
  JHEP {\bf 0208}, 040 (2002)
  [arXiv:hep-th/0204121].



\bibitem{Berkovits:2002ag}
  N.~Berkovits and V.~Pershin,
  ``Supersymmetric Born-Infeld from the pure spinor formalism of the open
  superstring,''
  JHEP {\bf 0301}, 023 (2003)
  [arXiv:hep-th/0205154].


\bibitem{Berkovits:2000ph}
  N.~Berkovits and B.~C.~Vallilo,
  ``Consistency of super-Poincare covariant superstring tree amplitudes,''
  JHEP {\bf 0007}, 015 (2000)
  [arXiv:hep-th/0004171].








\bibitem{Berkovits:2006bk}
  N.~Berkovits and C.~R.~Mafra,
  ``Some superstring amplitude computations with the non-minimal pure spinor
  formalism,''
  JHEP {\bf 0611}, 079 (2006)
  [arXiv:hep-th/0607187].



\bibitem{Aisaka:2002sd}
  Y.~Aisaka and Y.~Kazama,
  ``A new first class algebra, homological perturbation and extension of pure
  spinor formalism for superstring,''
  JHEP {\bf 0302}, 017 (2003)
  [arXiv:hep-th/0212316]
;




\bibitem{Grassi:2004xr}
  P.~A.~Grassi and P.~Vanhove,
  ``Topological M theory from pure spinor formalism,''
  Adv.\ Theor.\ Math.\ Phys.\  {\bf 9}, 285 (2005)
  [arXiv:hep-th/0411167]
;
  ``Higher-loop amplitudes in the non-minimal pure spinor formalism,''
  JHEP {\bf 0905}, 089 (2009)
  [arXiv:0903.3903 [hep-th]].











\bibitem{Policastro:2006vt}
  G.~Policastro and D.~Tsimpis,
  ``R**4, purified,''
  Class.\ Quant.\ Grav.\  {\bf 23}, 4753 (2006)
  [arXiv:hep-th/0603165].




\bibitem{Alexandrov:2007pd}
  V.~Alexandrov, D.~Krotov, A.~Losev and V.~Lysov,
  ``On Pure Spinor Superfield Formalism,''
  JHEP {\bf 0710}, 074 (2007)
  [arXiv:0705.2191 [hep-th]].





\bibitem{Mafra:2009wq}
  C.~R.~Mafra,
  ``Superstring Scattering Amplitudes with the Pure Spinor Formalism,''
  arXiv:0902.1552 [hep-th].




\bibitem{Gerigk:2009va}
  S.~Gerigk and I.~Kirsch,
  ``On the Relation between Hybrid and Pure Spinor String Theory,''
  JHEP {\bf 1003}, 106 (2010)
  [arXiv:0912.2347 [hep-th]].




\bibitem{Gomez:2010ad}
  H.~Gomez and C.~R.~Mafra,
  ``The Overall Coefficient of the Two-loop Superstring Amplitude Using Pure
  Spinors,''
  arXiv:1003.0678 [hep-th].














\bibitem{Gomez:2009qd}
  H.~Gomez,
  ``One-loop Superstring Amplitude From Integrals on Pure Spinors Space,''
  arXiv:0910.3405 [Unknown].







\bibitem{Aisaka:2009yp}
  Y.~Aisaka and N.~Berkovits,
  ``Pure Spinor Vertex Operators in Siegel Gauge and Loop Amplitude
  Regularization,''
  JHEP {\bf 0907}, 062 (2009)
  [arXiv:0903.3443 [hep-th]].





\bibitem{Berkovits:2006vi}
  N.~Berkovits and N.~Nekrasov,
  ``Multiloop superstring amplitudes from non-minimal pure spinor formalism,''
  JHEP {\bf 0612}, 029 (2006)
  [arXiv:hep-th/0609012].





\bibitem{Hoogeveen:2007tu}
  J.~Hoogeveen and K.~Skenderis,
  ``BRST quantization of the pure spinor superstring,''
  JHEP {\bf 0711}, 081 (2007)
  [arXiv:0710.2598 [hep-th]]
;
  ``Decoupling of unphysical states in the minimal pure spinor formalism I,''
  JHEP {\bf 1001} (2010) 041
  [arXiv:0906.3368 [hep-th]].




















































\bibitem{Candelas:1984yd}
  P.~Candelas and D.~J.~Raine,
  ``Spontaneous Compactification And Supersymmetry In D = 11 Supergravity,''
  Nucl.\ Phys.\  B {\bf 248}, 415 (1984).



























































































































\end{thebibliography}
\end{document}